\documentclass[preprint,showpacs,preprintnumbers,nofootinbib,amsmath,amssymb,aps,floatfix]{revtex4}

\usepackage{dcolumn}
\usepackage{graphicx}
\usepackage[dvips]{color}

\def\LL{\left\langle}	
\def\RR{\right\rangle}	
\def\LP{\left(}		
\def\RP{\right)}	
\def\PAR#1#2{ {\frac{\partial #1}{\partial #2}} }

\newcommand{\BE}{\begin{displaymath}}
\newcommand{\EE}{\end{displaymath}}
\newcommand{\BNE}{\begin{equation}}
\newcommand{\ENE}{\end{equation}}
\newcommand{\BEA}{\begin{eqnarray}}
\newcommand{\EEA}{\nonumber\end{eqnarray}}

\newcommand{\ie}{{\em i.e.,}}
\newcommand{\eg}{{\em e.g.,}}

\newcommand{\Tr}{{\rm Tr}}

\newcommand{\Dslash}{\makebox[0pt][l]{\,/}D}
\newcommand{\cO}{{\cal O}}

\def\figref#1{Fig.~\ref{fig:#1}}

\def\figrefs#1#2{Figs.~\ref{fig:#1} and \ref{fig:#2}}

\def\chpt{\raise0.4ex\hbox{$\chi$}PT}
\def\schpt{S\raise0.4ex\hbox{$\chi$}PT}
\def\rschpt{rS\raise0.4ex\hbox{$\chi$}PT}
\def\aschpt{ASHM\raise0.4ex\hbox{$\chi$}PT}
\newcommand{\vslash}{\ensuremath{v\!\!\! /}}
\def\eqn#1{\label{eq:#1}}
\def\eq#1{Eq.~(\ref{eq:#1})}

\def\tabref#1{Table~\ref{tab:#1}}

\begin{document}

\title{\bf\large Lattice QCD ensembles with four flavors of highly improved staggered quarks}

\author{A. Bazavov}
\affiliation{Physics Department, Brookhaven National Laboratory, Upton, NY 11973, USA}
\author{C. Bernard}
\affiliation{Department of Physics, Washington University, St. Louis, MO 63130, USA}
\author{C. DeTar}
\affiliation{Department of Physics and Astronomy, University of Utah, Salt Lake City, UT 84112, USA}
\author{W. Freeman}
\affiliation{Department of Physics, The George Washington University, Washington, DC 20052, USA}
\author{Steven Gottlieb}
\affiliation{Department of Physics, Indiana University, Bloomington, IN 47405, USA}
\author{U.M. Heller}
\affiliation{American Physical Society, One Research Road, Ridge, NY 11961, USA}
\author{J.E. Hetrick}
\affiliation{Physics Department, University of the Pacific, Stockton, CA 95211, USA}
\author{J. Komijani}
\affiliation{Department of Physics, Washington University, St. Louis, MO 63130, USA}
\author{J. Laiho}
\affiliation{Department of Physics and Astronomy, University of Glasgow, Glasgow G12 8GG, Scotland, UK}
\author{L. Levkova\footnote{present address: Department of Physics, University of Arizona, Tucson,
AZ 85721, USA}}
\affiliation{Department of Physics and Astronomy, University of Utah, Salt Lake City, UT 84112, USA}
\author{J. Osborn}
\affiliation{Argonne Leadership Computing Facility, Argonne National Laboratory, Argonne, IL
60439, USA}
\author{R.L. Sugar}
\affiliation{Department of Physics, University of California, Santa Barbara, CA 93106, USA}
\author{D. Toussaint}
\affiliation{Department of Physics, University of Arizona, Tucson, AZ 85721, USA}
\author{R.S. Van de Water}
\affiliation{Theoretical Physics Department, Fermi National Accelerator Laboratory, Batavia 60510, USA}
\author{Ran Zhou}
\affiliation{Department of Physics, Indiana University, Bloomington, IN 47405, USA}
\author{[MILC Collaboration]}

\date{\today}

\begin{abstract}
We present results from our simulations of quantum chromodynamics (QCD)
with four flavors of quarks: $u$, $d$, $s$, and $c$. These simulations are
performed with a one-loop Symanzik improved gauge action,
and the highly improved staggered quark (HISQ) action.
We are generating gauge configurations
with four values of the lattice spacing ranging from 0.06~fm to 0.15~fm,
and three values of the light quark mass, including the value for which
the Goldstone pion mass is equal to the physical pion mass. We discuss
simulation algorithms, scale setting, taste symmetry breaking, and
the autocorrelations of various quantities. We also present  results
for the topological susceptibility which demonstrate the improvement
of the HISQ configurations relative to those generated earlier
with the asqtad improved staggered action.

\end{abstract}

\maketitle
\section{Introduction}
\label{INTRO}

Over the past decade, we have generated a large library of gauge configuration
ensembles with three flavors of improved staggered (asqtad) quarks~\cite{RMP}.
These ensembles are publicly available, and they are being used by our
group and several others to study a wide variety of problems in high-energy
and nuclear physics~\cite{RMP}. A number of the most challenging problems 
that we and other lattice gauge theorists are pursuing, however, require
a level of precision that is beyond the reach of the current asqtad
ensembles, and generating additional ones with smaller lattice spacings
and lighter up and down quark masses would be very computationally expensive.
We have therefore begun generating a new library of gauge configuration
ensembles using the highly improved staggered quark (HISQ) action introduced
by the HPQCD Collaboration~\cite{hisq2003,hisq2004,HISQ}. In this paper, we 
describe the ensembles produced to date, and report on the initial calculations 
performed with them.

The HPQCD Collaboration developed the HISQ action to reduce the
taste-symmetry violations associated with staggered quarks, and to
improve the quark dispersion relation sufficiently so that charm quarks
can be simulated at lattice spacings accessible with
today's computers.  They tested the new action using HISQ valence quarks on
quenched gauge configurations and on ones we had generated with
asqtad sea quarks~\cite{hisq2003,hisq2004,HISQ}. More recently,
they have obtained impressive results for charm and heavy-light physics 
again using HISQ valence quarks on asqtad 
configurations~\cite{HPQCD_APPS,HPQCD_CHARM,Mc,HPQCD_DC2,HPQCD_B,HPQCD_FD2,HPQCD_MASSRATIO,HPQCD_FD3,HPQCD_SEMI11}.
We have performed tests of scaling in the lattice spacing using HISQ
valence quarks with  gauge configurations generated
with HISQ sea quarks~\cite{MILC_TESTS}.
We found that lattice artifacts for the HISQ action are
reduced by approximately a factor of three from those of the
asqtad action for the same lattice spacing, and taste splittings in the
pion masses are reduced sufficiently to enable us to undertake simulations
with the mass of the Goldstone pion at or near the physical pion
mass. Moreover, the improvement in the quark dispersion relation
enables us to include charm sea quarks in the simulations.

\newcommand{\on}{\phantom{1}}
\newcommand{\col}{\phantom{:}}
\newcommand{\hun}{\phantom{101}}
\begin{table}[t]
\begin{center}
\begin{tabular}{|c|c|c|c|c|c|}
\hline 
$\approx a$ (fm) & $m_l/m_s$ &$N_s^3\times N_t$  &\ $M_\pi L$\ \ & $M_\pi$ (MeV) 
&\ $N_{\rm lats}$\  \\
\hline
0.15  & 1/5  & $16^3\times 48$ & 3.78 & 306.9(5)  & 1021 \\
0.15 & 1/10 & $24^3\times 48$ & 3.99 & 214.5(2)  & 1000 \\
0.15  & 1/27 & $32^3\times 48$ & 3.30 & 131.0(1) & 1020 \\
\hline
0.12 & 1/5  & $24^3\times 64$ & 4.54 & 305.3(4)  &  1040\\
0.12  & 1/10 & $24^3\times 64$ & 3.22 & 218.1(4)  & 1020 \\
0.12  & 1/10 & $32^3\times 64$ & 4.29 & 216.9(2)  & 1000\\
0.12  & 1/10 & $40^3\times 64$ & 5.36 & 217.0(2) & 1029 \\
0.12 & 1/27 & $48^3\times 64$ & 3.88 &  131.7(1) & 1000\\ 
\hline
0.09  & 1/5  & $32^3\times 96$ & 4.50& 312.7(6) & 1011\\
0.09  & 1/10 & $48^3\times 96$ &  4.71 &  220.3(2) & 1000\\
0.09  & 1/27 & $64^3\times 96$ & 3.66 & 128.2(1) & 235+467 \\
\hline
0.06  & 1/5  & $48^3\times 144$ & 4.51 & 319.3(5) & 1000\\
0.06  & 1/10 & $64^3\times 144$ & 4.25 & 229.2(4) & \on 435+227\\
0.06  & 1/27 & $96^3\times 192$ & 3.95 & 135.5(2) & \on 240 \\
\hline
\end{tabular}
\vspace{0.2in}
\caption{HISQ gauge configuration ensembles with strange
and charm quark masses set at or very close to their physical values.
The first column gives the lattice spacing for which we were aiming,
which in all cases turned out to be a good approximation to the actual
lattice spacing that could only be determined after the lattices were created.
The second column gives the ratio of the simulation mass of the light
quark to the physical mass of the strange quark, the third the lattice
dimensions, the fourth the product of the Goldstone pion mass and the spatial
extent of the lattice, and the fifth the Goldstone pion mass in MeV.
The pion masses were converted to physical units using the $f_{p4s}$ scale
setting described in section~\protect\ref{SCALE}.
The quoted errors include only
the statistical errors on the pion mass and $f_{p4s}$ in lattice units
in the individual ensemble; they do not include systematic errors such as the errors on
the physical values of $f_{p4s}$ in Table~\protect\ref{fp4s_values_table}.
The sixth column gives the number of equilibrated gauge configurations.
Where the sixth column is the sum of two numbers, these are the numbers
of lattices generated with the RHMC and RHMD algorithms respectively,
as discussed in section~\protect\ref{SIMULATIONS}.
We plan to save approximately 1,000 configurations
in each ensemble, so those for which $N_{lats}\geq 1,000$ are considered 
to be complete. 
\label{physical_strange}
}
\end{center}
\end{table}

\begin{table}[t]
\begin{center}
\begin{tabular}{|c|c|c|c|c|}
\hline
$m_{l1}/m_s$ & $m_{l2}/m_s$ & $m_s'/m_s$ & $N_s^3\times N_t$&\ $N_{\rm lats}$\  \\
\hline
0.10 & 0.10 & 0.10 & $32^3\times 64$ & 1020 \\
0.10  & 0.10 & 0.25 & $32^3\times 64$ & 1020\\
0.10  & 0.10  & 0.45 & $32^3\times 64$  & 1020\\
0.10 & 0.10 & 0.60 & $32^3\times 64$ & 1020\\
0.25  & 0.25 & 0.25 & $24^3\times 64$ & 1020\\
0.20 & 0.20 & 0.60 & $24^3\times 64$ & 1020\\
\col 0.175&\on 0.175&0.45 & $32^3\times 64$  & 1020\\
0.10  & 0.25 & 0.45 & $32^3\times 64$ & 1020 \\
\hline
\end{tabular}
\vspace{0.2in}
\caption{HISQ gauge configuration ensembles with lighter-than-physical
strange quark masses. 
All ensembles have a lattice spacing of $a\approx 0.12$~fm and charm-quark
mass as close as possible to its physical value. The first
two columns give the ratio of the light quark masses to the physical strange
quark mass. (We distinguish between the masses of the two light quarks because
in the ensemble in the last row they are different.) 
The third column gives
the ratio of the simulation strange-quark mass to the physical strange-quark
mass, and the fourth column shows the lattice dimensions. The fifth
shows the number of equilibrated configurations.
\label{unphysical_strange}
}
\end{center}
\end{table}

Given the successes of the HISQ action, we have embarked on a project
to generate ensembles of gauge configurations using it along with a
one-loop Symanzik improved gauge action~\cite{symanzik_action}.
We are working at four different lattice spacings, $a\approx 0.06$,
0.09, 0.12 and 0.15~fm, in order to control extrapolations to the
continuum limit. We include up, down, strange, and charm sea quarks.
For most ensembles, the masses of the strange and charm quarks,
$m_s$ and $m_c$, respectively, are fixed at their physical values.
For these ensembles, the up and down quark
masses are taken to be degenerate with a common mass $m_l$, 
which has a negligible effect ($< 1\%$) on isospin-averaged quantities.
We are generating configurations with three values of the light quark
mass: $m_l=m_s/5$, $m_s/10$, and the value such that the Goldstone pion
mass is as close as possible to the physical pion mass, which is
approximately $m_s/27$. 
Table~\ref{physical_strange} shows the current state of these ensembles.
Prior to the simulations, the lattice spacing and the physical values of
the quark masses can only be estimated. Their precise values are
outputs from the analysis described later in this paper.
Note that we have generated
three ensembles with $a\approx 0.12$~fm and
$m_l=m_s/10$ that differ only in their spatial volumes. The
purpose of using three different volumes is to enable tests of finite-size
effects. The factor governing such effects, $e^{-M_\pi L}$,
varies by a factor of eight over this range of spatial sizes, so we
expect to have a sufficient lever arm for these tests.
In Table~\ref{tab:finitesize}, we compare the values of the plaquette, 
the strange and light quark condensates, and $r_1/a$ on these three lattices.
A comparison of finite size effects for the pion and kaon masses
and leptonic decay constants on these configurations with the predictions
of chiral perturbation theory can be found in Ref.~\cite{fKfpi}.

\begin{table}[h]
\begin{tabular}{|c|l|l|l|}
\hline
                 & \ \ \ $L=24$       &\ \ \ $L=32$       & \ \ \ $L=40$   \\
\hline
Plaquette        & 0.556621( 7) & 0.556631( 5) & 0.556623( 3) \\
\hline
$\bar\psi\psi_l$ & 0.016957(33) & 0.017020(18) & 0.017068(13) \\
\hline
$\bar\psi\psi_s$ & 0.061764(18) & 0.061719(12) & 0.061735( 8) \\
\hline
$r_1/a$          & 2.585(19)    & 2.626(13)    & 2.611( 9)     \\
\hline
\end{tabular}
\caption{ \label{tab:finitesize} 
Effects of spatial size in the $a\approx 0.12$~fm $m_l=m_s/5$
ensembles, which differ only in the spatial size of the lattices.
$\bar\psi\psi$ is in units of the lattice spacing, and is shown for both the
light quark and strange quark masses.
Values for $r_1/a$ come from a fit to the heavy-quark
potential over spatial separations
from 1.5 to 4.0, and time lengths 5 to 7.
}
\end{table}

With the HISQ action, as with less improved staggered fermion actions,
each lattice fermion species corresponds to four ``tastes'' of fermions
in the continuum limit. To eliminate the three unwanted tastes from
the quark sea, we use the fourth-root
procedure for each of the sea-quark flavors, up, down, strange, and
charm. For numerical and theoretical arguments justifying this 
fourth-root procedure we refer the
reader to Refs. \cite{Fourth_root_num} and \cite{Fourth_root_theory}.

We have also generated a limited number of ensembles with the strange-quark
mass lighter than its physical value, because including such ensembles has
proven very useful in controlling chiral extrapolations of physical quantities.
In one of those ensembles, we also chose different values for the two
light-quark masses, up and down, to probe for isospin-breaking effects.
These ensembles are listed in Table~\ref{unphysical_strange}.

We note that even though we are generating some ensembles with the Goldstone
pion mass at the physical value and with the strange quark mass near its 
physical value, controlling the chiral expansion using a variety of 
other ensembles with different quark masses is still very useful for 
several reasons: (1) The lattice spacing and the physical values
of the quark masses can only be estimated prior to the simulations.
Their precise values are outputs from a detailed analysis made after the
ensembles have been created in which the unphysical light- and strange-quark
mass ensembles play important roles.
(2) We can both determine the dependence on the 
strange sea-quark mass and extrapolate to the limit where that mass 
vanishes, which is important for studies of a variety of physical quantities. 
(3) Since the continuum and chiral expansions are closely connected 
through staggered chiral perturbation theory, controlling the chiral 
expansion helps us to take the continuum limit with small 
extrapolation errors. (4) Statistical fluctuations of physical quantities 
with up and down sea and valence masses at their physical values tend 
to be larger than at somewhat higher masses, so including higher-mass
ensembles in a 
chiral fit can significantly decrease the final errors. (5) Having a 
range of strange and light sea-quark masses allows us to determine 
low-energy constants of the chiral expansion, which are important fundamental
parameters of QCD in their own right.

The details of our simulations and their present status are given in
Sec.~\ref{SIMULATIONS}. In this section, we also compare 
configurations generated with the RHMC and RHMD algorithms. Although
we have used the former for nearly all the ensembles produced to date,
at the smallest lattice spacings the gauge fields appear to be smooth
enough to use the latter, which leads to a significant savings in
computer time.
Scale setting is addressed in Sec.~\ref{SCALE}, where we give results based
both on the Sommer parameter $r_1$~\cite{Sommer:1993ce,MILC_r1}
and on the decay constants of (fictitious)
pseudoscalar mesons~\cite{HPQCD_SCALE}.
In Sec.~\ref{TOPO}, we present results for
the topological susceptibility of the QCD vacuum obtained from the
HISQ ensembles. Tunneling between different topological sectors appears
to be sufficient for them to equilibrate. The topological susceptibility 
provides a particularly 
stringent test of the HISQ gauge configurations because it is computed 
from them without involving valence quarks. Thus, comparison of the topological 
susceptibility on the asqtad and HISQ configurations directly demonstrates
the improvement in the gauge configurations.
In Sec.~\ref{TASTE}, we discuss
taste-symmetry violations in the light-light, heavy-light and heavy-heavy 
pseudoscalar sectors, and demonstrate the improvement in the light-light
sector provided by the HISQ action,
while in Sec.~\ref{AUTO}, we 
discuss autocorrelation times for a variety of physical variables.
In Sec.~\ref{CONCL}, we present our conclusions, and in an appendix, we
tabulate the taste-symmetry breaking pseudoscalar meson mass splittings,
which can be used in the staggered chiral perturbation theory analysis of
masses and matrix elements on these ensembles.

\section{Simulations}
\label{SIMULATIONS}

The details of our simulations were given in Ref.~\cite{MILC_TESTS}. Here
we briefly outline our approach in order to make this paper self-contained. 
We are using a
one-loop Symanzik improved gauge action~\cite{symanzik_action} and the
HISQ action~\cite{hisq2003,hisq2004,HISQ}. The gauge action includes
$1\times 1$ and $1\times 2$ planar Wilson loops, and a $1\times 1 \times 1$
``parallelogram''. The coefficients of these terms are calculated
perturbatively,  and are then tadpole improved. They include both 
one-gluon-loop and one-quark-loop contributions~\cite{QL-COEFF}. 
The HISQ action consists of a Fat7 smearing of the gauge links,
then a projection of each smeared link onto a unitary matrix, followed
by an ``asq'' smearing with twice the Lepage term and including the Naik
term, a third nearest-neighbor coupling. This is designed to ensure 
that the quark action is order $a^2$ improved.
The use of two levels of smearing
produces a smooth gauge field seen by the quarks, which leads to 
the reduction in taste-symmetry violations mentioned previously. Finally, 
a modification of the Naik term~\cite{HISQ} improves the dispersion 
relation of the charm quark sufficiently that it can be included in our 
simulations.

We have used the rational hybrid Monte Carlo (RHMC) algorithm~\cite{RHMC} to
generate configurations with two exceptions discussed below. We use
different molecular dynamics step sizes for the gauge and fermion
parts of the action, with three gauge steps for each fermion 
one~\cite{SEXTON_WEINGARTEN}.  The Omelyan integration algorithm 
is employed for both the gauge and fermion parts of the 
action~\cite{SEXTON_WEINGARTEN,OMELYAN}, and five pseudofermion
fields are used, each with a rational function approximation for the fractional
powers of the fermion determinants. The projection of links onto U(3)
after the Fat7 smearing can lead to spikes in the fermion force,
which give rise to low acceptance rates, especially on coarse lattices.
These spikes are smoothed out by means of a guiding Hamiltonian in
the molecular dynamics evolution~\cite{MILC_TESTS}, and the algorithm
is made exact by means of a Metropolis accept/reject step at the end of each
trajectory, which employs the exact Hamiltonian.   

The simulation parameters for the ensembles with physical strange-quark
mass are given in Table~\ref{physical_strange_params}, and those for
ensembles with lighter-than-physical strange quarks in 
Table~\ref{unphysical_strange_params}.

\begin{table}[t]
\setlength{\tabcolsep}{0.8mm}
\rule{0.0in}{0.0in}\hspace{-0.2in}
\begin{tabular}{|c|lll|l|l|lclc|}
\hline
$10/g^2$ & $am_l$ & $am_s$ & $am_c$ &\ \ \ \ $u_0$ &\ \ \ $\epsilon_N$ 
& $s$ & len. & \ \ \ $\epsilon$       & acc. \\ 
\hline
%
%
5.80 &  0.013 & 0.065  & 0.838& 0.85535 & -0.3582* & 5  & 1.0 & 0.033 & 0.73 \\

5.80 &  0.0064 & 0.064  & 0.828 & 0.85535 & -0.3484 &  5& 1.0 & 0.020 & 0.71 \\

5.80 &  0.00235 & 0.0647  & 0.831 & 0.85535 & -0.3503 & 5 & 0.5& 0.018 & 0.68 \\
\hline

6.00 & 0.0102 & 0.0509 & 0.635 & 0.86372 & -0.2308* & 5 & 1.0& 0.036 & 0.66 \\

6.00 & 0.00507 & 0.0507 & 0.628 & 0.86372 & -0.2248 & 5& 1.0 & 0.033 & 0.68 \\

6.00 & 0.00507 & 0.0507 & 0.628 & 0.86372 & -0.2248 & 5 & 1.0 & 0.025 & 0.64 \\

6.00 & 0.00507 & 0.0507 & 0.628 & 0.86372 & -0.2248 & 5 & 1.0 & 0.014 & 0.66 \\

6.00 & 0.00184 & 0.0507 & 0.628  & 0.86372 & -0.2248 & 5 & 1.0 & 0.0091& 0.64 \\
\hline

6.30 & 0.0074 & 0.037  & 0.440  & 0.874164& -0.1204 & 6 & 1.5  & 0.031 & 0.67 \\

6.30 & 0.00363 & 0.0363  & 0.430  &0.874164& -0.1152 & 6 & 1.5 & 0.0214 & 0.66 \\

6.30 & 0.0012 & 0.0363  & 0.432   & 0.874164& -0.1162 &6  & 1.5& 0.0115 & 0.67 \\
\hline

6.72& 0.0048 & 0.024  & 0.286  & 0.885773& -0.0533 & 6 & 2.0 & 0.02 & 0.74 \\

6.72 & 0.0024 & 0.024  & 0.286 & 0.885773& -0.0533 & 6 & 2.0 & 0.0167 & 0.76 \\

6.72 & 0.00084 & 0.0231  & 0.274  & 0.885773& -0.0491 & 6 & 2.0 & 0.0125 & n.a. \\

\hline

%
%
\end{tabular}
\vspace{0.1in}
\caption{ \label{physical_strange_params}
Parameters of the HISQ ensembles with the strange and charm quark masses
at or  close to their physical values. The ensembles are listed in the same 
order as in Table~\ref{physical_strange}. The first column gives the gauge
coupling constant $10/g^2$, and the second, third and fourth columns the masses
of the light, strange and charm quarks in lattice units. The
fifth column is the tadpole coefficient $u_0$ obtained from the fourth-root
of the plaquette, and the sixth is
the mass-dependent correction to the tree-level improvement of the charm
quark dispersion relation, or Naik term, $\epsilon_N$. (The two values of 
$\epsilon_N$ with stars use the bare mass, rather than the tree-level  mass in
Eq.~(26) of Ref.~\cite{HISQ}, giving rise to a difference that appears at
order $am_c^6$).  $s$ is the separation of stored configurations and len., 
the length of a trajectory, both in simulation time units. 
$\epsilon$ is the molecular dynamics step size, and acc.  is the fraction of 
the trajectories accepted. 
}
\end{table}

\begin{table}[t]
\setlength{\tabcolsep}{0.8mm}
\rule{0.0in}{0.0in}\hspace{-0.2in}
\begin{tabular}{|llll|lc|}
\hline
\ \ $am_{l1}$ &\ \  $am_{l2}$ &\ \  $am_s$ &\  $am_c$ & \ \ \ $\epsilon$ & acc. \\
\hline
0.00507   & 0.00507   & 0.00507  & 0.628  & 0.02   & 0.67  \\ 

0.00507   & 0.00507   & 0.012675 & 0.628  & 0.02   & 0.68  \\ 

0.00507   & 0.00507   & 0.022815 & 0.628  & 0.02   & 0.70  \\ 

0.00507   & 0.00507   & 0.0304   & 0.628  & 0.0227 & 0.68  \\ 

0.01275   & 0.01275   & 0.01275  & 0.640  & 0.0278 & 0.71  \\ 

0.0102    & 0.0102    & 0.03054  & 0.635  & 0.0294 & 0.74  \\ 

0.0088725 & 0.0088725 & 0.022815 & 0.628  & 0.02   & 0.70  \\ 

0.00507   & 0.012675  & 0.022815 & 0.628  & 0.02   & 0.73  \\ 

\hline
\end{tabular}
\vspace{0.1in}
\caption{\label{unphysical_strange_params}
Parameters of the HISQ ensemble with lighter-than-physical 
strange quarks. The ensembles are listed in the same 
order as in Table~\ref{unphysical_strange}, and the notation is the same
as in Table~\ref{physical_strange_params},
except that we distinguish between the masses of the two light quarks, 
because for the ensemble in the last row they are unequal. For all of these 
ensembles, $10/g^2=6.00$, $u_0=0.86372$, $\epsilon_N=-0.2248$, $s=5$, 
and len.=1.0.
}
\end{table}

\newcommand{\pbp}{\bar\psi\psi}
\newcommand{\plaq}{\mathrm{Plaq}}
\renewcommand{\Dslash}{\makebox[0pt][l]{\null\,\big /}D}

Obtaining a good acceptance rate with the RHMC algorithm requires a step size
small enough that the change in the action over a trajectory is of order one, 
and on large lattices this requires a small simulation time step.  However, 
this requirement is unphysical in the sense that we are interested in
intensive quantities, not extensive ones. 
In our earlier series of simulations with the asqtad quark 
action~\cite{RMP}, several of the largest ensembles were run with the 
RHMD algorithm in which the accept-reject step at the end of each trajectory 
is omitted.  This is an attractive possibility because it allows 
running at a step size limited by its effects on physical quantities, and 
because the lattice generation can be done in single precision.
(Double precision is necessary for the RHMC algorithm on large lattices 
because the numerical error in evaluating the action at the beginning
and end of the trajectory in order to compute its change, $\Delta S$, becomes
of order one, making the accept/reject step nonsensical.
The cost of double precision can be reduced by running
the conjugate gradient in the molecular dynamics steps in single precision,
followed by
double precision refinement of the result.) 

\begin{figure}[t]
\hspace{-0.1in}
\begin{tabular}{lll}
\includegraphics[width=0.33\textwidth]{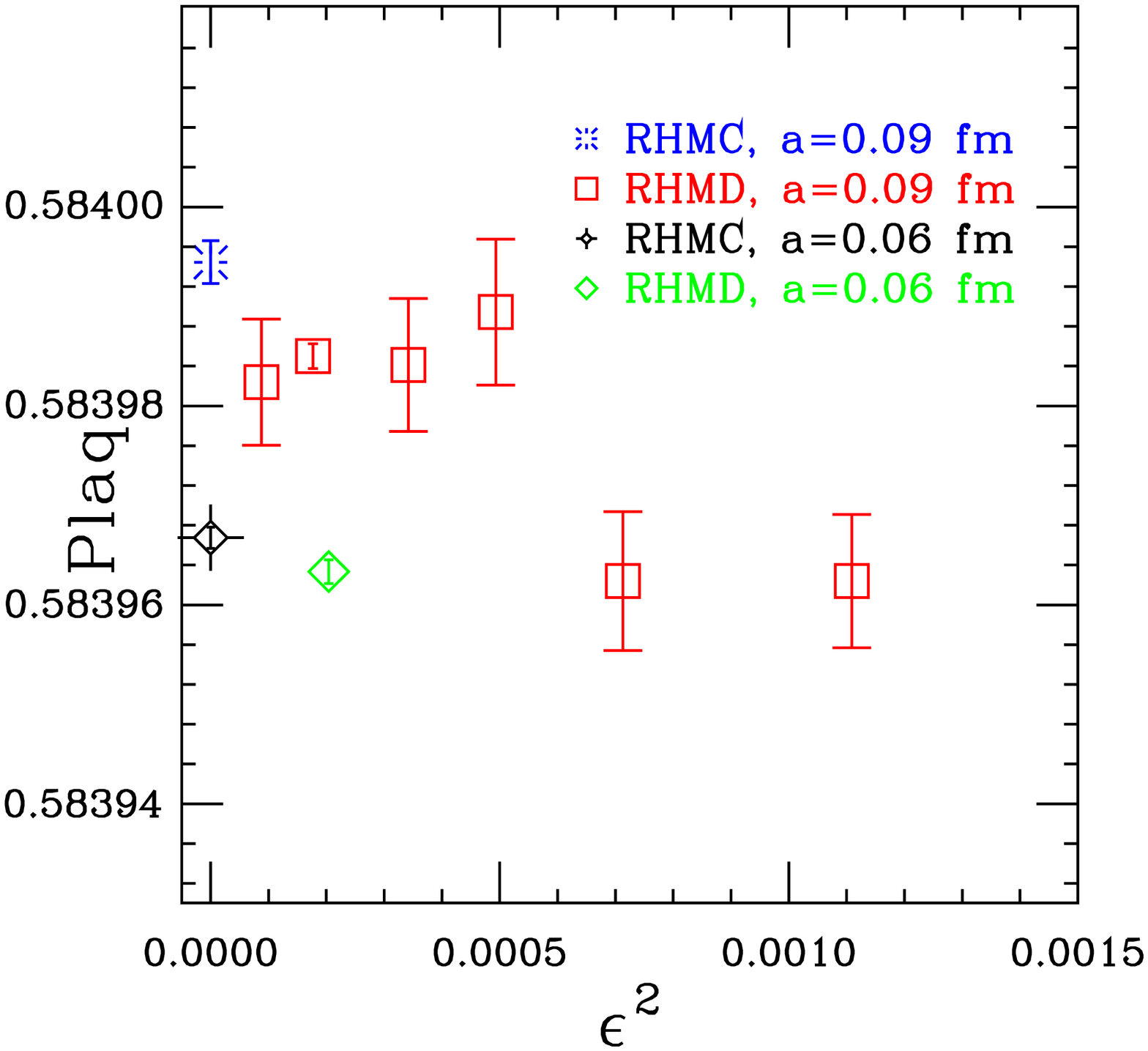} &
\includegraphics[width=0.33\textwidth]{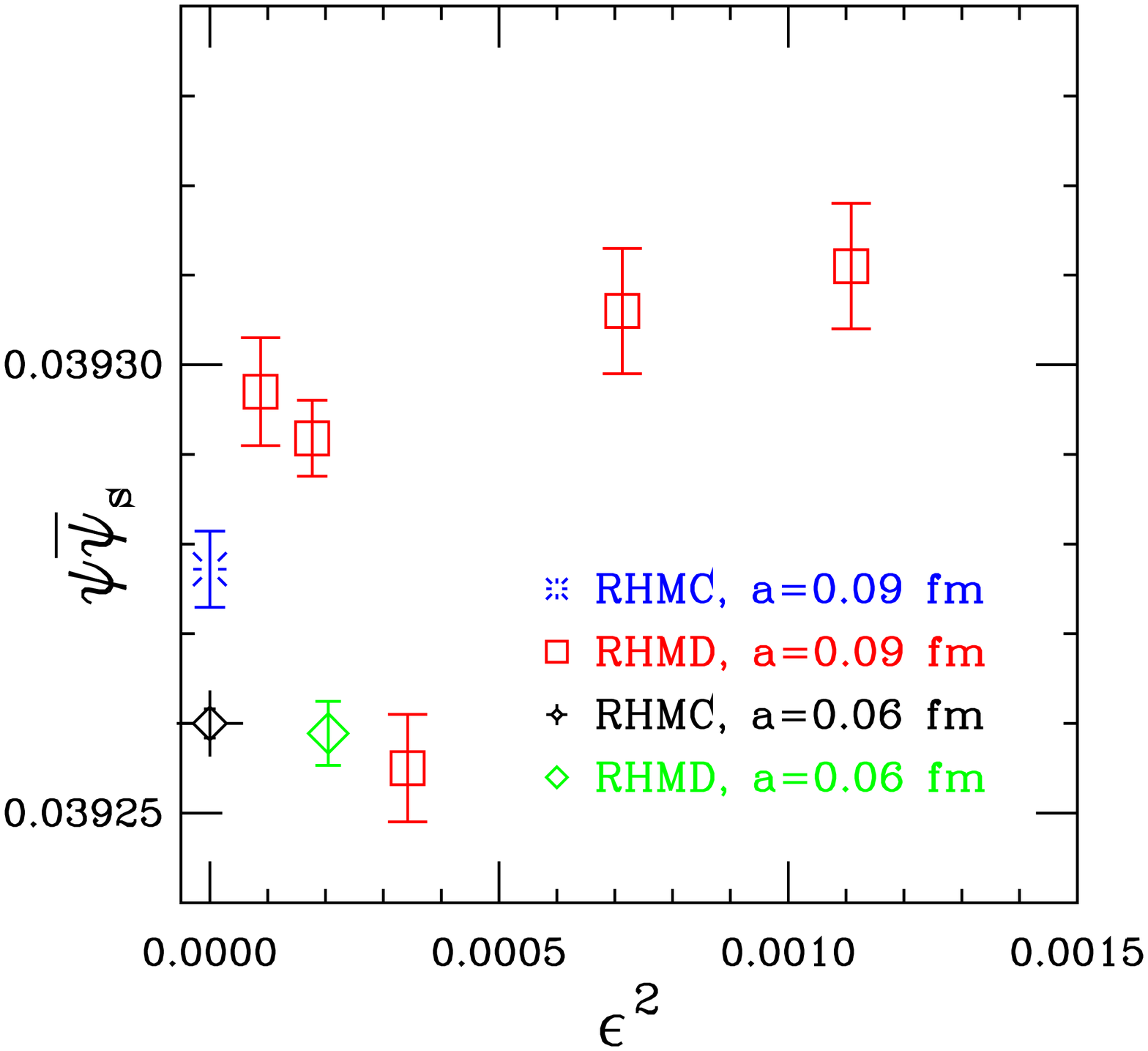} &
\includegraphics[width=0.33\textwidth]{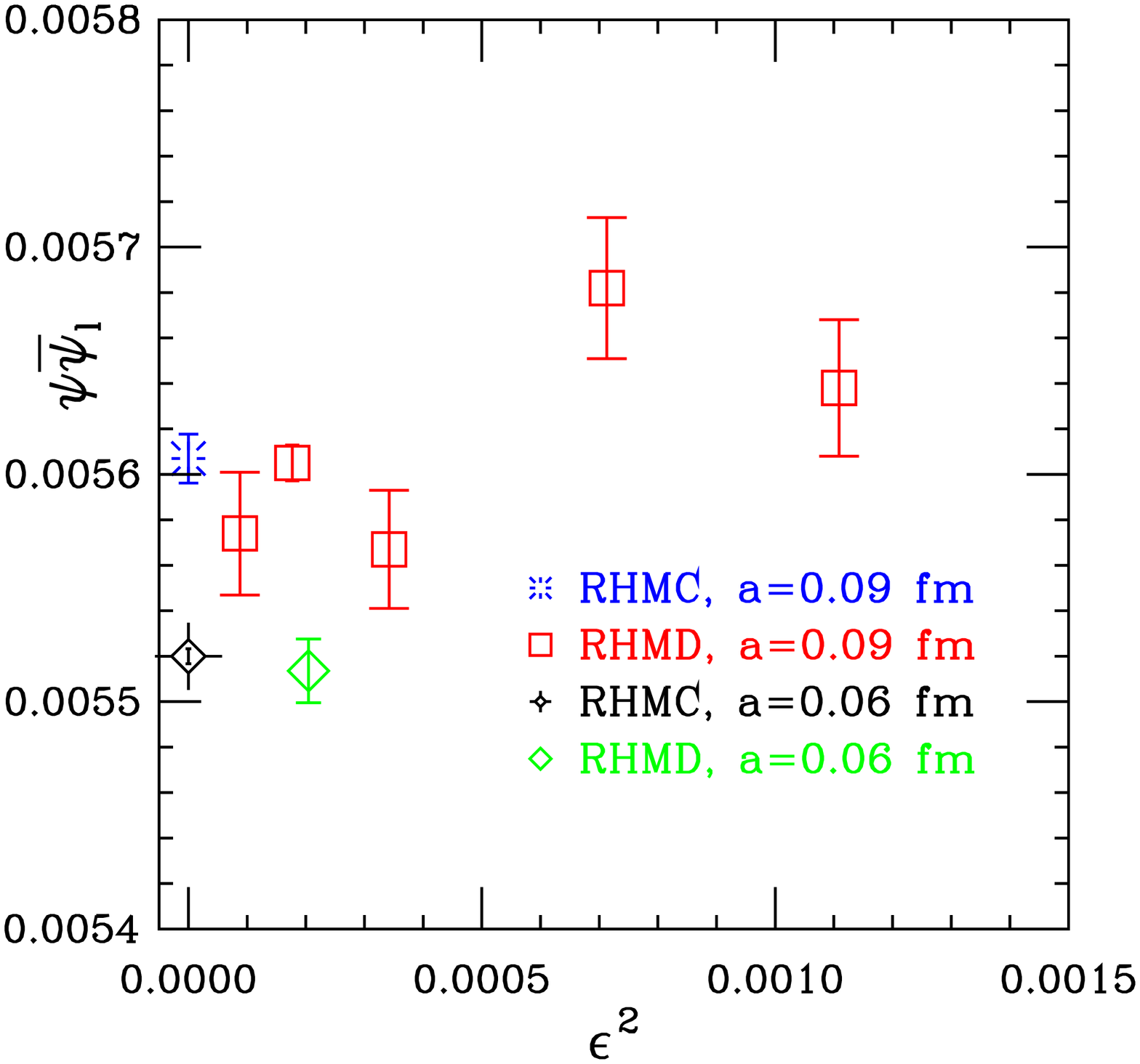} \\
\end{tabular}
\caption{Comparison of the RHMC and RHMD algorithms for the
$a\approx 0.09$~fm, physical quark mass and $a\approx 0.06$~fm,
$m_l=m_s/10$ ensembles. The RHMD points are plotted at the
value of the molecular dynamics step size $\epsilon$ for which they were 
generated, and are given by
red squares for the $a\approx 0.09$~fm ensemble, and by
green diamonds for the $a\approx 0.06$~fm one. 
The RHMC points are plotted at $\epsilon=0$, and are given by blue
bursts for the $a\approx 0.09$~fm ensemble and by fancy black
diamonds for the $a\approx 0.06$~fm one.
The left panel shows the plaquette,
the central panel the strange-quark $\langle\pbp\rangle$,
and the right panel the light-quark $\langle\pbp\rangle$, 
all as a function of $\epsilon^2$. A significant fraction of the total
sample for $a\approx 0.09$~fm was run at $\epsilon^2=0.000177$, which 
is why the error bars at that point are so small.
Both the light and strange quark $\langle\pbp\rangle$ are given in 
lattice units, and both are calculated in double precision.
The points for the $a\approx 0.06$~fm ensemble have been
shifted vertically to move them into the range of the graph,
the plaquette downward by 0.031643, the strange and light quark
$\langle\pbp\rangle$ upward by 0.0157497 and 0.00185230, respectively.
\label{fig:rhmd_all3}
}
\end{figure}

It is not possible to use the RHMD algorithm on coarse lattices because of
the occurrence of spikes in the fermion force. For such lattices, if
one uses the RHMD algorithm with the exact Hamiltonian in the molecular 
dynamics evolution of the
gauge fields, then it is necessary to use a very small step size in order
to perform the integration of the molecular dynamics equations accurately,
whereas if one uses the guiding Hamiltonian, which ignores these spikes,
there will be a significant deviation of the resulting gauge fields from
the correct ones. However, as the lattice spacing decreases, the gauge
fields become smoother, and spikes in the fermion force become less
pronounced and very infrequent. This is fortunate, since it is the very
challenging ensembles with small lattice spacings for which the RHMD
algorithm would provide the greatest gain if applicable. We have tested
the efficacy of the RHMD algorithm on two of our ensembles, those with
lattice spacing $a\approx 0.09$~fm and physical light quark mass, and 
with lattice spacing $a\approx 0.06$~fm and $m_l=m_s/10$. For the first
of these ensembles we have run a few RHMD trajectories at a number of 
different step sizes, and generated a significant fraction of our total 
sample with this algorithm at a step size of $\epsilon=0.0133$, which
compares with the step size of 0.0115 used in the RHMC part of the ensemble.
Although the step size used in the RHMD sub-ensemble is only a little larger
than that in the RHMC sub-ensemble, large gains are realized from running
in single precision and from not rejecting up to one third of the trajectories.

We have compared several quantities computed  separately on the RHMC and 
RHMD sub-ensembles.
Results for the plaquette and the light- and strange-quark condensates
$\langle\pbp\rangle$ 
on the $a\approx 0.09$~fm physical quark mass ensemble
are shown in
Fig.~\ref{fig:rhmd_all3}.  The plaquette shows the expected effect
proportional to $\epsilon^2$, while the strange-quark condensate is
less regular, and the light quark condensate has large errors.  The
points with small error bars at $\epsilon^2=(0.0133)^2\approx 0.000177$ 
(third point from the left)
are the points where 
production running was done with the RHMD algorithm.  To estimate the 
physical significance of
these effects, we may suppose that the effect is largely a change in the 
lattice spacing,
\BNE \Delta a = \PAR{a}{\plaq}\PAR{\plaq}{\epsilon^2} \epsilon^2 \ \ \ .\ENE
Using the difference of the plaquettes at $\epsilon=0$ and $\epsilon=0.0133$,
and the differences in plaquettes among ensembles at $a\approx 0.12$, $0.09$ 
and $0.06$ fm, we find that this corresponds to a shift in the lattice 
spacing of
about $1.4\times10^{-5}$ fm, or a fractional shift of about $1.6\times 10^{-4}$.

\begin{table}
\begin{tabular}{|l|l|l|l|}
\hline
         & \ \ RHMC & \ \ RHMD  &\ difference \\
\hline
$aM_\pi$       & $0.05716(9)$ & $0.05717(5)$ & $0.00001(11)$ \\
$aM_{\bar s s}$ & $0.30608(14)$ & $0.30624(10)$ & $0.00016(17)$ \\
$aM_{\eta_c}$   & $1.32708(6)$     & $1.32718(5)$ & $0.00010(8)$ \\
\hline
\end{tabular}
\vspace{0.1in}
\caption{
\label{tab:rhmdmasses}
Pseudoscalar meson masses for the $a\approx 0.09$~fm physical quark mass 
RHMC and RHMD ($\epsilon=0.0133$)
sub-ensembles.  The fit ranges and fit forms are not necessarily those that 
would be used for a final spectrum analysis, but care has been taken to 
use the same fit ranges in each sub-ensemble.
}
\end{table}

The black fancy diamonds and green diamonds 
in Fig.~\ref{fig:rhmd_all3} show the
plaquette and strange $\langle\pbp\rangle$ for the $a\approx 0.06$~fm 
$m_l=m_s/10$ run
at $\epsilon=0$ (RHMC) and $\epsilon=0.0143$ (RHMD), with an arbitrary constant
added to put them in the range of this graph.  Here the effects are somewhat 
smaller than in the $a\approx 0.09$ fm ensemble,
likely because the 
gauge configurations are smoother at the smaller lattice spacing.

While the plaquette and $\langle\pbp_s\rangle$ are determined accurately enough so that
the step size effects are visible, it is much more interesting to see how 
physical quantities are affected by the step size errors.  Unfortunately, 
the only such quantities that can be determined with the required precision 
are the pseudoscalar meson masses.  Table~\ref{tab:rhmdmasses} shows 
masses for the light-light, strange-strange and charm-charm pseudoscalar 
mesons in the RHMC and RHMD sub-ensembles.

From these results, we see that the use of the RHMD algorithm introduces
a small systematic error, in comparison to other uncertainties.
For all but a few quantities, this error is also smaller than
the statistical error.  Among our current projects,
the only quantity for which we cannot use the RHMD results is 
the interaction measure
in high temperature QCD, where a delicate subtraction of the plaquette
and $\langle\pbp\rangle$ between the hot and cold lattices is needed.
With this one exception, we believe that it is safe to use the RHMD
algorithm for $a\leq 0.09$~fm, although we did use RHMC for the
$a\approx 0.09$~fm, $m_l=m_s/5$ and $m_s/10$ ensembles.

\section{Setting the lattice scale}
\label{SCALE}

Since lattice computations produce quantities in units of the lattice
spacing, converting to physical units, such as MeV or fm, requires accurate
knowledge of the lattice spacing.  This is done by computing some physical
quantity which is experimentally well known and can be accurately computed
on the lattice.  Examples in common use include the $\Omega^-$ mass, the
splittings of charmonium levels, and the pion or kaon decay constant.  
These quantities
can be used to fix the lattice spacing in physical units, and thus to
convert other measurements to physical units.  In simulation programs
with multiple ensembles, it is often convenient to use an intermediate
interpolating quantity which is easily measured on the lattice, but may
not be an experimentally accessible quantity, such as the commonly used
Sommer scales $r_0$ and $r_1$~\cite{MILC_r1}.  
This interpolating quantity can be
fixed by calculating a physical quantity on some of the ensembles,
often with a fitting or smoothing procedure, or even from a completely
different set of simulations.

\begin{table}
\begin{tabular}{|c|c|c|l|}
\hline
$m_l/m_s$   & $f_{p4s}$ (MeV) & $M_{p4s}$ (MeV) &\phantom{XXX}   ratio  
\\
\hline
1/5    & 157.7(0.1)(0.8)     & 437.6(0.5)(2.0)    & 0.3604(5)(30) \\
\hline
1/10    & 155.5(0.2)(0.6)    & 435.1(0.3)(1.1)    & 0.3575(6)(18) \\
\hline
1/27    & 154.0(0.4)(0.6)    & 434.5(0.1)(0.6)    & 0.3544(10)(15) \\
\hline
\end{tabular}
\caption{\label{fp4s_values_table} 
The decay constant $f_{p4s}$, the meson mass $M_{p4s}$ and their ratio 
from the 2+1 flavor asqtad analysis.
The numbers in parentheses are the statistical and systematic errors, in
that order. 
}
\end{table}

Strictly speaking, observables measured on the lattice take on their
physical values only in the continuum limit at physical valence- and
sea-quark masses.
Therefore, the definition of the lattice spacing for unphysical quark masses
or non-zero lattice spacings is somewhat arbitrary. The definition can be
made by requiring that the interpolating quantity be independent of the
lattice spacing and sea-quark masses, or by fixing its dependence on these
quantities to some plausible ansatz, such as that predicted by chiral 
perturbation theory at some order. Of course, different ans\"atze 
are expected to give identical results only in the continuum, physical 
quark-mass limit.

Here we present determinations of the lattice scale on the HISQ ensembles
from two such interpolating quantities,
$r_1$ and $f_{p4s}$.  The first of these is calculated from
the static-quark potential, and was used extensively in our earlier
simulation program using the asqtad action as described in Ref.~\cite{RMP}.
The second quantity, $f_{p4s}$, is a variant of a method suggested 
in Ref.~\cite{HPQCD_SCALE}.  This is the decay constant of a pseudoscalar
meson with valence-quark masses $0.4$ times the strange-quark mass.
Ultimately, we expect that our preferred scale-setting will be based
on $f_\pi$. The scale setting via $f_\pi$ will largely follow the same
approach as described for $f_{p4s}$ here, since $f_\pi$ and $f_{p4s}$
are determined from the same types of correlators and differ only in the
quark masses.
The value $0.4\,m_s$ was chosen to be heavy enough to allow relatively
cheap determination on the lattice ensembles, but light enough that
chiral perturbation theory can accurately describe its dependence
on the valence- and sea-quark masses.

\begin{table}[t]
\setlength{\tabcolsep}{0.8mm}
\rule{0.0in}{0.0in}\hspace{-0.2in}
\begin{tabular}{|c|lll|lcc|}
\hline
$10/g^2$ & $am_l$ & $am_s$ & $am_c$ & $r_1/a$ & $af_{p4s}$&
$am_{p4s}$ \\ 
\hline
5.80 &  0.013 & 0.065  & 0.838&
2.059(23) & 0.12150(18) & 0.02744(9)\\

5.80 &  0.0064 & 0.064  & 0.828 &
2.073(13) & 0.12042(11) & 0.02744(5) \\

5.80 &  0.00235 & 0.0647  & 0.831 &
2.089(8) & 0.11948(6) & 0.02762(3) \\
\hline

6.00 & 0.0102 & 0.0509 & 0.635 &
2.575(17) & 0.09780(12) & 0.02139(6) \\

6.00 & 0.00507 & 0.0507 & 0.628 &
2.585(19) & 0.09614(14) & 0.02111(7) \\

6.00 & 0.00507 & 0.0507 & 0.628 &
2.626(13)  & 0.09613(9) & 0.02118(5)  \\

6.00 & 0.00507 & 0.0507 & 0.628 &
2.614(9)  & 0.09605(7) & 0.02113(4) \\

6.00 & 0.00184 & 0.0507 & 0.628  &
2.608(8)  & 0.09530(5) & 0.02130(2)\\
\hline

6.30 & 0.0074 & 0.037  & 0.440  &
3.499(24)  & 0.07093(11) & 0.01482(5)\\

6.30 & 0.00363 & 0.0363  & 0.430  &
3.566(14)  & 0.06953(6) & 0.01467(3)\\

6.30 & 0.0012 & 0.0363  & 0.432   &
3.565(13)  & 0.06865(4) & 0.01462(2)\\
\hline

6.72& 0.0048 & 0.024  & 0.286  &
5.342(16)  & 0.04660(7) & 0.00918(3) \\

6.72 & 0.0024 & 0.024  & 0.286 &
5.376(14)  & 0.04545(5) & 0.00896(2) \\
\hline
\end{tabular}
\vspace{0.1in}
\caption{\label{physical_strange_params_r1_fp4s}
$r_1/a$, $af_{p4s}$ and $am_{p4s}$, measured on the ensembles
with physical strange- and charm-quark masses. These quantities are used to 
determine the lattice spacing, which is given in the next table.
(Note that $m_{p4s}$ is the quark mass corresponding to $f_{p4s}$.)
}
\end{table}

Since we do not know the correct valence strange-quark mass until
after the lattice spacing is fixed, $f_{p4s}$ and $0.4\,m_s$ must
be determined self-consistently.  Roughly speaking, this is done
by finding the valence-quark mass $a m_q$ where $ a f_{p4s}$
and $a M_{p4s}$, the mass of the pseudoscalar meson with valence
quark mass $0.4\,m_s$, have their expected ratio,
as we now explain.

\begin{table}[t]
\setlength{\tabcolsep}{0.8mm}
\rule{0.0in}{0.0in}\hspace{-0.2in}
\begin{tabular}{|c|lll|ll|ll|ll|}
\hline
$10/g^2$ & $am_l$ & $am_s$ & $am_c$ & $a_{r_1}$ (fm) & $a_{f_{p4s}}$ (fm) &
 $am_s^{r_1,tuned}$ & $am_s^{f_{p4s},tuned}$ & $am_c^{r_1,tuned}$ & $am_c^{f_{p4s},tuned}$ \\
\hline
5.80 & 0.013   & 0.065  & 0.838 & 0.1510(20) & 0.1520(8) &
 0.0668(15) & 0.0678(3) & 0.8472(16) & 0.8514(16) \\
5.80 & 0.0064  & 0.064  & 0.828 & 0.1499(14) & 0.1528(6) &
 0.0671(3) & 0.0686(1) & 0.8407(7) & 0.8489(7)\\
5.80 & 0.00235 & 0.0647 & 0.831 & 0.1488(11) & 0.1531(7) &
 0.0655(5) & 0.0690(1) & 0.8351(9) & 0.8522(9) \\
\hline
6.00 & 0.0102  & 0.0509 & 0.635 & 0.1207(11) & 0.1224(6) &
 0.0516(7) & 0.0535(2) & 0.6363(9) & 0.6489(10) \\
6.00 & 0.00507 & 0.0507 & 0.628 & 0.1202(12) & 0.1220(5) &
 0.0514(7) & 0.0528(2) & 0.6310(10) & 0.6375(10) \\
6.00 & 0.00507 & 0.0507 & 0.628 & 0.1184(10) & 0.1220(5) &
 0.0500(5) & 0.0530(1) & 0.6241(8) & 0.6376(8) \\
6.00 & 0.00507 & 0.0507 & 0.628 & 0.1189(9) & 0.1219(5) &
 0.0504(4) & 0.0528(1) & 0.6263(7) & 0.6372(7) \\
6.00 & 0.00184 & 0.0507 & 0.628 & 0.1191(7) & 0.1221(6) &
 0.0507(3) & 0.0533(1) & 0.6271(10) & 0.6378(10) \\
\hline
6.30 & 0.0074  & 0.037  & 0.440 & 0.0888(8) & 0.0887(5) &
 0.0367(5) & 0.0370(2) & 0.4396(7) & 0.4393(7) \\
6.30 & 0.00363 & 0.0363 & 0.430 & 0.0872(7) & 0.0882(4) &
 0.0358(3) & 0.0367(1) & 0.4288(5) & 0.4325(5) \\
6.30 & 0.0012  & 0.0363 & 0.432 & 0.0871(6) & 0.0880(4) &
 0.0357(3) & 0.0365(1) & 0.4294(5) & 0.4323(5) \\
\hline
6.72 & 0.0048 & 0.024   & 0.286 & 0.0582(4) & 0.0583(3) &
 0.0225(1) & 0.0230(1) & 0.2768(3) & 0.2772(3) \\
6.72 & 0.0024 & 0.024   & 0.286 & 0.0578(4) & 0.0577(2) &
 0.0224(1) & 0.0224(1) & 0.2750(2) & 0.2755(3) \\
\hline
\end{tabular}
\vspace{0.1in}
\caption{ \label{physical_strange_lat_scale}
Lattice spacing and the retuned strange- and charm-quark masses,
set by using $r_1$ and $f_{p4s}$ quantities for ensembles with 
physical strange-and charm-quark masses.
For the $am_l=0.00184$ ensemble the $J/\Psi$ mass is not available,
so the $\eta_c$ mass was used in tuning instead.
}
\end{table}

For now, we use ``physical'' values of $f_{p4s}$ and $M_{p4s}$ determined
from fits to pseudoscalar meson masses and amplitudes in the MILC
2+1 flavor asqtad action ensembles~\cite{RMP}.
In practice, these come from evaluating the chiral-continuum extrapolation 
fit function for the pseudoscalar meson mass and decay constant at the 
appropriate bare-quark masses:  strange sea-quark $m_s = m_s^{\rm physical}$, 
light-to-strange sea-quark mass ratio $m_l / m_s$ equal to the simulated 
value for that ensemble, and degenerate valence-quark masses 
$m_{\rm valence} = 0.4\, m_s^{\rm physical}$.

These values are shown in Table~\ref{fp4s_values_table}.
Clearly, when a similar analysis of the HISQ pseudoscalar mesons
is available, $f_{p4s}$ and $m_{p4s}$ from that analysis will
be used instead. 

The determination of $a f_{p4s}$ on one ensemble begins with calculating 
pseudoscalar
meson masses and decay constants for several valence-quark masses.
In our ensembles, the valence-quark masses used include $0.3\,m_s^\prime$,
$0.4\,m_s^\prime$ and $0.6\,m_s^\prime$, where $m_s^\prime$ is the 
strange-quark mass estimated before the ensemble was started, which in 
most cases
is the same as the strange sea-quark mass used in generating the
ensemble.  We then 
interpolate the square $\LP M_{p4s}/f_{p4s} \RP^2$ as a quadratic function of
the bare valence-quark mass.
We use $\LP M_{p4s}/f_{p4s} \RP^2$ rather
than alternatives, such as $f_{p4s}/M_{p4s}$, since we expect $M_{p4s}^2$ to
be approximately linear in $am_q$, and $f_{p4s}$ to be approximately
constant.  The valence-quark mass at which $(M_{p4s}/f_{p4s})^2$
has the desired value is then the physical $0.4\,m_s$.   
The decay constant in lattice units $af_{p4s}$ is then found
by 
similarly interpolating $(af_{p4s})^2$ as a quadratic function of the 
bare quark masses and evaluating it at $0.4\,m_s$.
Next, the lattice spacing is obtained by requiring that $f_{p4s}$
on this ensemble equals the desired value.
Finally, $M_{p4s}$ is found from $f_{p4s}$ and the ratio.
Errors on this quantity are found by a jackknife analysis, where
blocks of 16 lattices are omitted from the completed ensembles,
and blocks of 8 lattices from the ensembles that are approximately
50\% completed.
The resulting values for $a f_{p4s}$, $a m_{p4s}$ and the lattice spacing 
in fermi are shown in Tables~\ref{physical_strange_params_r1_fp4s},  
\ref{physical_strange_lat_scale}, and \ref{unphysical_strange_params_r1_fp4s}.
In these tables,
the error in $a_{r_1}$~($a_{f_{p4s}}$)~(fm) 
includes the statistical and systematic
errors in the physical value of $r_1$ ($f_{p4s}$), combined in quadrature
with the statistical error in $r_1/a$ ($a f_{p4s}$).
For the physical value of $r_1$ we use $r_1=0.3106(8)(14)(4)$, obtained from
an analysis of pseudoscalar decay constants on lattices generated with
the asqtad quark action~\cite{ASQTAD_R1}.  Errors in the physical lattice
spacings in these tables combine the errors in the physical value of $r_1$ with
the statistical errors in $r_1/a$.
Again, we will eventually determine the physical $r_1$ with the HISQ ensembles. 
Because $r_1$ is not experimentally observable, however, we will still need 
to use a measured quantity such as $f_\pi$ or the $\Upsilon$ 1S-2S mass 
splitting to obtain $r_1$ in physical units.


\begin{table}[t]
\setlength{\tabcolsep}{0.8mm}
\rule{0.0in}{0.0in}\hspace{-0.2in}
\begin{tabular}{|llll|lcc|ll|}
\hline
\ \ $am_{l1}$ &\ \  $am_{l2}$ &\ \  $am_s$ &\  $am_c$ &\ \ \ $r_1/a$ & $af_{p4s}$&
$am_{p4s}$ & $a_{r_1}$ (fm) & $a_{f_{p4s}}$ (fm) \\
\hline
0.00507   & 0.00507   & 0.00507  & 0.628  & 2.675(16) & 0.08973(9) & 0.01900(5) &
 0.1161(10) & 0.1139(10) \\

0.00507   & 0.00507   & 0.012675 & 0.628  & 2.676(16) & 0.09109(9) & 0.01944(4) &
0.1161(9) & 0.1156(5)  \\

0.00507   & 0.00507   & 0.022815 & 0.628  & 2.653(13) & 0.08973(10) & 0.01903(4) &
 0.1171(8) & 0.1139(5)  \\

0.00507   & 0.00507   & 0.0304   & 0.628  & 2.647(15) & 0.09383(8) & 0.02042(4) &
 0.1173(9) & 0.1191(5) \\

0.01275   & 0.01275   & 0.01275  & 0.640  & 2.642(21) & 0.09391(14) & 0.02038(6) &
 0.1176(11) & 0.1192(5)  \\

0.0102    & 0.0102    & 0.03054  & 0.635  & 2.593(19) & 0.09569(15) & 0.02099(6) &
 0.1198(11) & 0.1214(5)  \\

0.0088725 & 0.0088725 & 0.022815 & 0.628  & 2.646(16) & 0.09416(12) & 0.02055(4) &
 0.1174(9) & 0.1195(5)  \\

0.00507   & 0.012675  & 0.022815 & 0.628  & 2.664(16) & ---      & ---       &
 0.1166(9) & \ \ \ \ ---\\

\hline
\end{tabular}
\vspace{0.1in}
\caption{\label{unphysical_strange_params_r1_fp4s}
$r_1/a$, $af_{p4s}$ and $am_{p4s}$ for the ensembles
with unphysical strange-quark mass. The lattice spacings determined
from these quantities are given in the last two columns.
Since the purpose of these lattices is to study dependence as the
quark masses are varied, and because we do not know the dependence
of the physical $f_{p4s}$ on the strange quark mass, the value
appropriate for $m_l=m_s/10$, 155.5(2)(6) MeV, is used for all of these 
ensembles. (As in Table~\ref{physical_strange_params_r1_fp4s}, $m_{4ps}$
is the quark mass corresponding to $f_{4ps}$).
}
\end{table}

To find the tuned charm-quark mass, we calculated pseudoscalar and vector
charmonium masses for two valence-quark masses: the sea charm-quark mass
and 0.9 times the sea charm-quark mass.  Using linear interpolation, we
find the valence charm quark mass where the spin-averaged charmonium
mass has its physical value.  Of course, this depends on the previous
determination of the lattice spacing, so we quote two values, where
the lattice spacing was determined from $r_1$ or from $f_{p4s}$.

In Table~\ref{physical_strange_lat_scale} we compare lattice spacings
and retuned strange- and charm-quark masses based on the $r_1$ and
$f_{p4s}$ scales. In general, we observe good agreement in the lattice
spacing and the strange-quark mass, with the largest difference in the
latter of about 5-6\% for some of the $a\approx 0.12$ and 0.15~fm ensembles.
As expected, the
difference in the lattice spacing and retuned quark masses decreases
towards the continuum and is small on the finest, $a\approx 0.06$~fm ensembles.

\section{Topological susceptibility}
\label{TOPO}

The topological susceptibility 
\begin{equation}
\chi_t = \langle \nu^2 \rangle/V
\label{eq:def_topo_susc}
\end{equation}
measures fluctuations in the total topological charge $\nu$ in the
space-time volume $V$.  The angle brackets represent an average over
the gauge-field configurations.  These fluctuations are suppressed at
small quark mass.  Leading-order continuum chiral perturbation theory
predicts \cite{Leutwyler:1992yt} that

\begin{equation}
  \chi_t V \approx x \equiv V\Sigma m^\prime,
\label{eq:chipt_topo_susc}
\end{equation}
where $\Sigma$ is the chiral condensate parameter and $m^\prime$ is
the reduced mass of the quarks:
\begin{equation}
  1/m^\prime = 1/m_u + 1/m_d  + 1/m_s + \ldots{} \, .
\end{equation}
The relation (\ref{eq:chipt_topo_susc}) is valid provided $x \gg 1$,
which is the case for our simulations.

For equal up- and down-quark masses we may use the
Gell-Mann-Oakes-Renner relation, also from leading-order chiral
perturbation theory, to rewrite this expression as
\begin{equation}
  f_\pi^2/(4 \chi_t) = 2/M_{\pi,I}^2 + 1/M_{ss,I}^2 + \ldots{},
 \label{eq:chiral}
\end{equation}
where $M^2_{ss,I} = 2M_{K,I}^2 - M_{\pi,I}^2$ is the squared mass of
the fictitious pseudoscalar meson containing two nonannihilating
quarks with masses equal to the strange quark.  In our
normalization the pion decay constant $f_\pi$ is approximately 130
MeV. In leading order {\it staggered} chiral perturbation theory, the
meson masses appearing in Eq.~(\ref{eq:chiral}) are
taste-singlet masses, as indicated by the subscript $I$.

\begin{table}[t]
\begin{tabular}{|c|c|c|c|}
\hline
$\approx a$~(fm)  &  $m_l/m_s$ & $M_{\pi,I}^2 r_0^2$ & \ \ \ $\chi_t r_0^4$ \\
  \hline
    0.12 & 1/5   & 1.006 & 0.0170(5) \\
\hline
    0.12 & 1/27 & 0.560 & 0.0108 (2) \\
\hline
    0.09 & 1/5   & 0.736 & 0.0124 (7) \\
\hline
    0.09 & 1/27 & 0.282 & 0.0059 (4) \\
\hline
    0.06 & 1/5   & 0.603 & 0.0075 (8) \\
\hline
  \end{tabular}
\caption{Topological susceptibility for the indicated HISQ ensembles
{\it vs.} taste-singlet pion mass squared in units of $r_0$. 
\label{tab:HISQtopo}
}
\end{table}

Confirming Eq.~(\ref{eq:chipt_topo_susc}) provides an essential test
of the fourth-root treatment of the fermion determinant, since it
gives a nearly direct measure of the influence of sea quarks on the
gauge field.  Since the right-hand side depends on the taste-singlet mass
splitting, it also tests the degree of improvement of the staggered
action.

In Ref.~\cite{ref:topo_susc2} we presented results for the topological
susceptibility over a wide range of light-quark masses and lattice
spacings with $2+1$ flavors of improved (asqtad) quarks.  Here, we use
the same methods to determine the topological susceptibility in the
presence of $2+1+1$ flavors of HISQ sea quarks and compare results with
the asqtad study.  We have studied five of the HISQ ensembles, namely
three with $m_l = m_s/5$ and $a \approx 0.06$, 0.09, and 0.12
fm, and two at the physical Goldstone pion mass with $m_l \approx m_s/27$
and $a \approx 0.09$ and 0.12~fm.

In brief, we use the Boulder discretization
\cite{DeGrand:1997gu,Hasenfratz:1998qk} of the topological charge
density
\begin{equation}
   \rho(x) = \frac{1}{32\pi^2} F^a_{\mu\nu} \widetilde F^a_{\mu\nu},
\end{equation}
measured after three HYP smoothing sweeps \cite{Hasenfratz:2001hp} of
the gauge field.  The topological susceptibility is then obtained from
density-density correlations:
\begin{equation}
  \chi_t = \int d^4x \, C(r) \ \ \ \mbox{with} \ \ \  C(r) = \langle \rho(x) \rho(0) \rangle \ ,
\label{eq:corrint}
\end{equation}
Although in the continuum limit this definition suffers from
ultraviolet singularities that require regularization
\cite{Luscher:2004fu}, such complications are unimportant at our
range of lattice spacings \cite{ref:topo_susc2}.

\begin{figure}[t]
\begin{tabular}{llll}
\vtop{
\hbox{ \includegraphics[height=.20\textwidth]{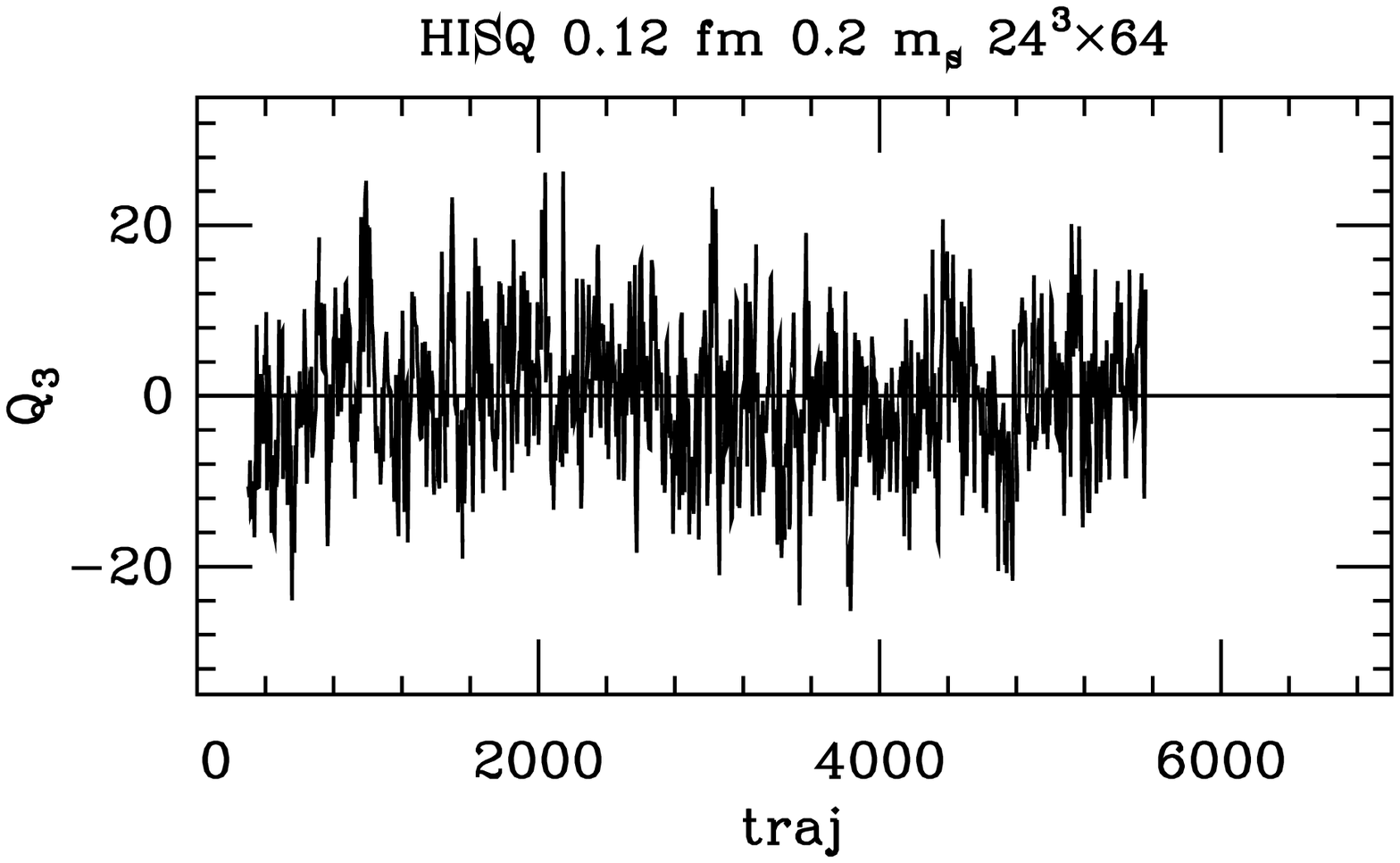}}}
&
\vtop{
\vspace{-1.16in}
\hbox{
\includegraphics[height=0.142\textwidth]{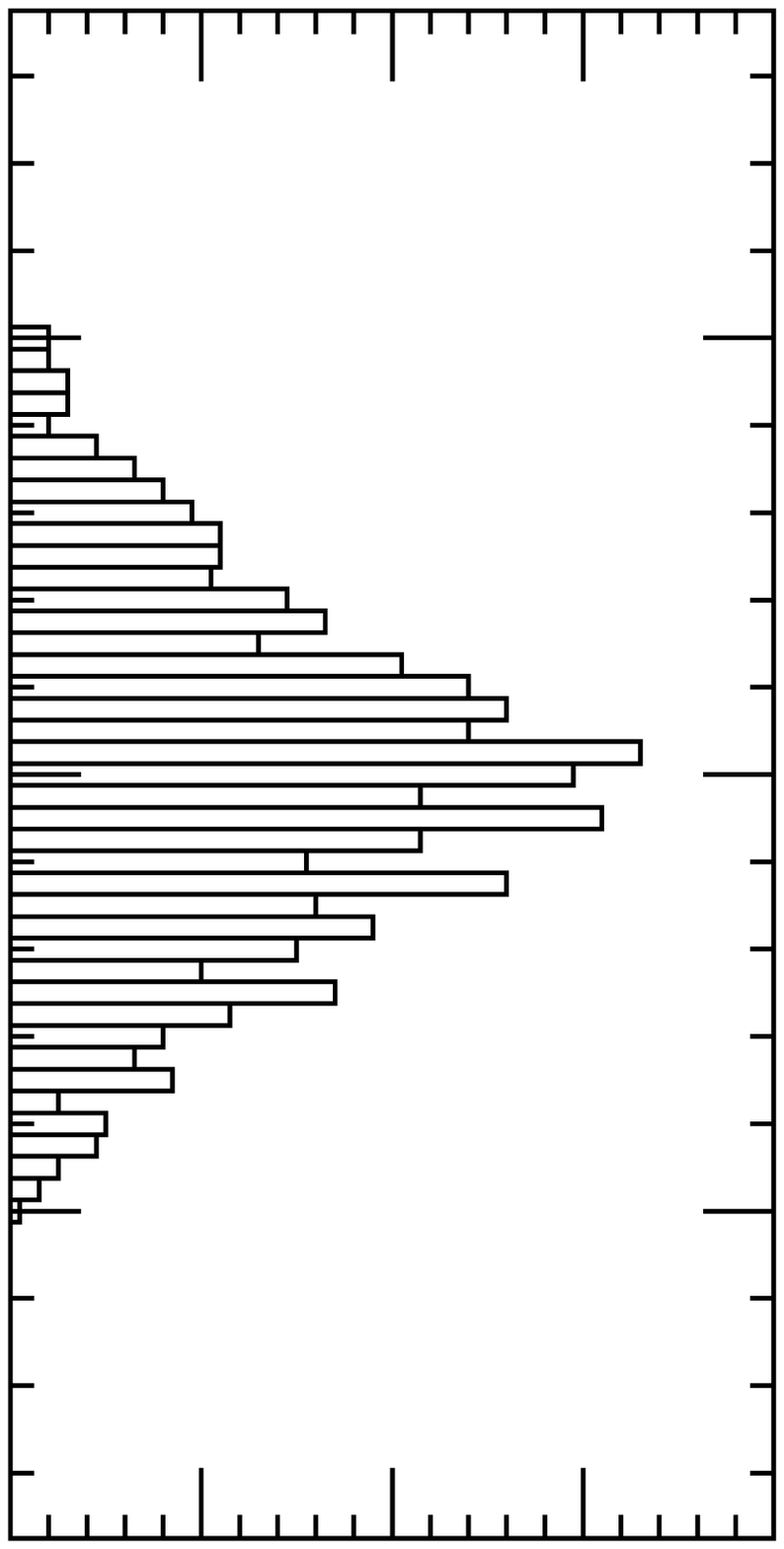}}}
&
\vtop{
\hbox{ \includegraphics[height=.20\textwidth]{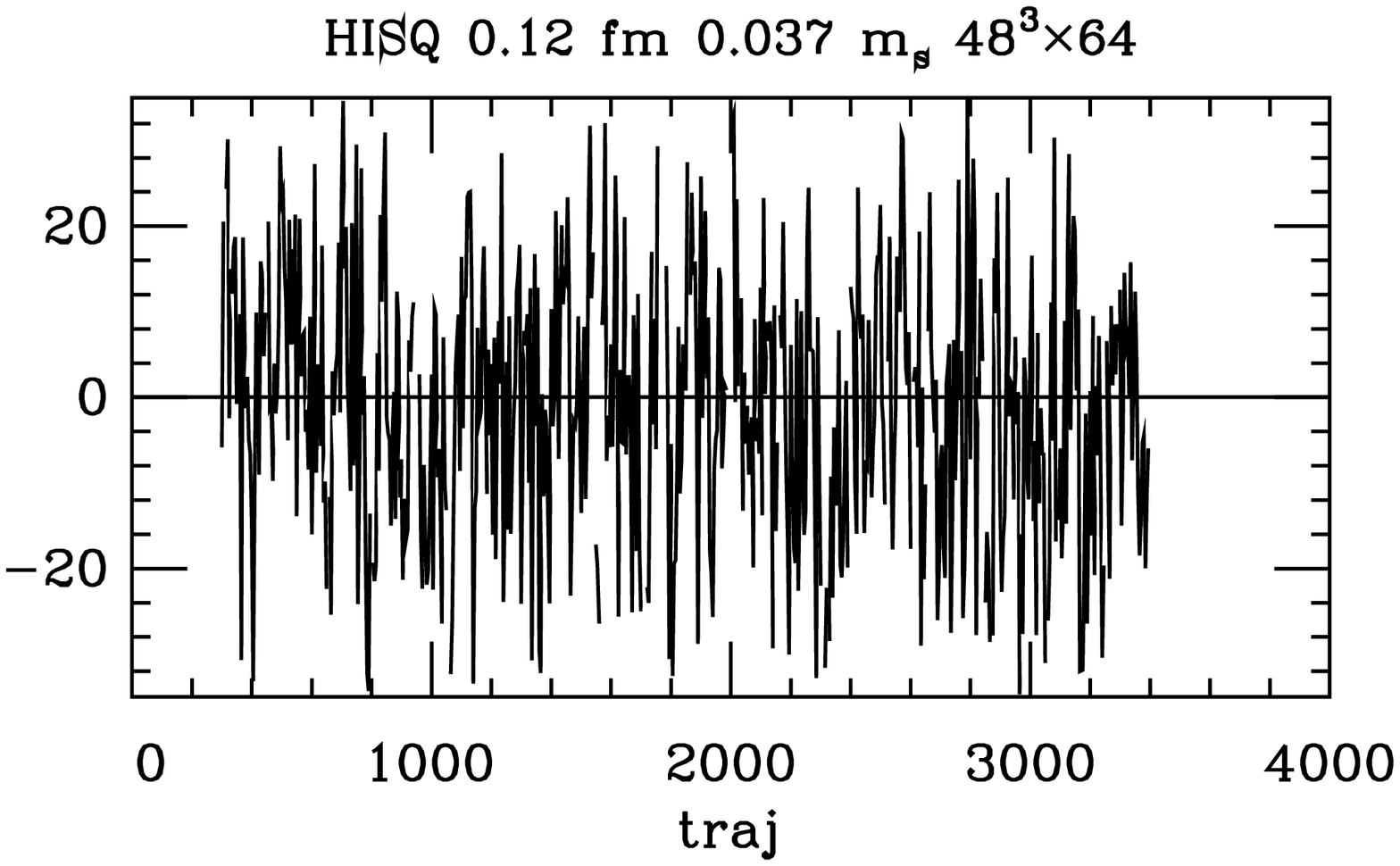}}}
&
\vtop{
\vspace{-1.16in}
\hbox{
\includegraphics[height=0.142\textwidth]{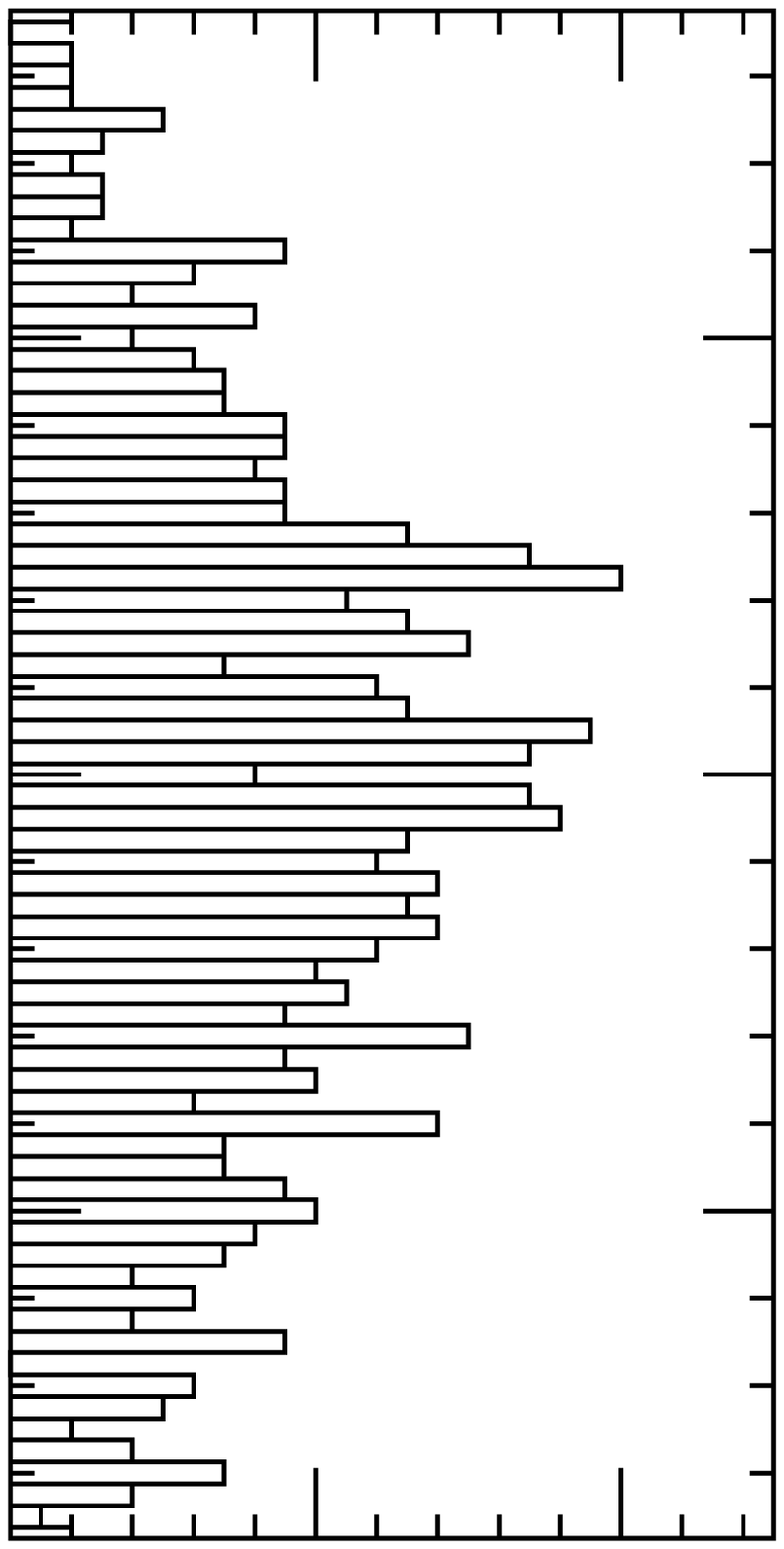}}}
\\
\vtop{
\hbox{ \includegraphics[height=.20\textwidth]{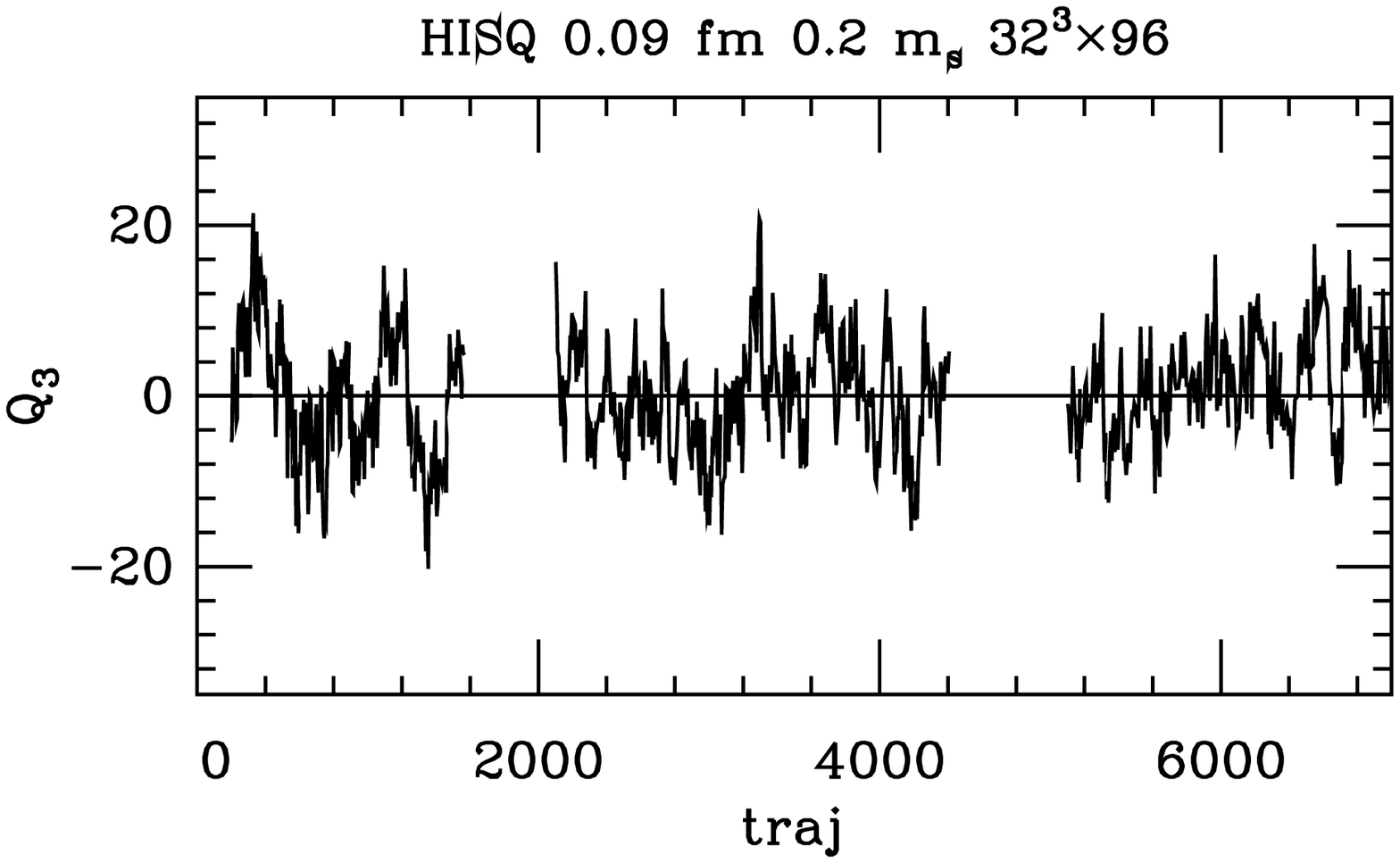}}}
&
\vtop{
\vspace{-1.16in}
\hbox{
\includegraphics[height=0.142\textwidth]{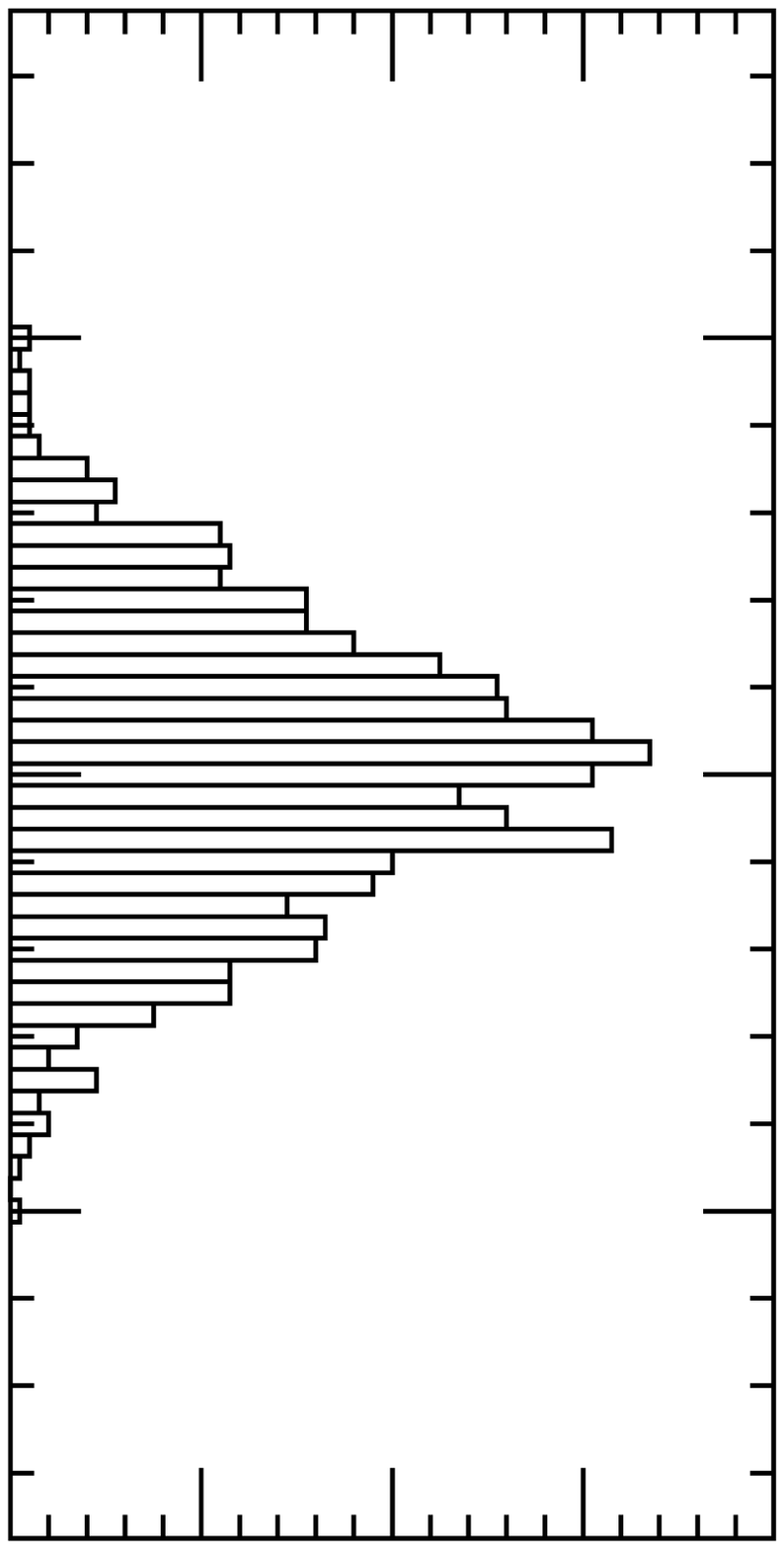}}}
&
\vtop{
\hbox{ \includegraphics[height=0.2\textwidth]{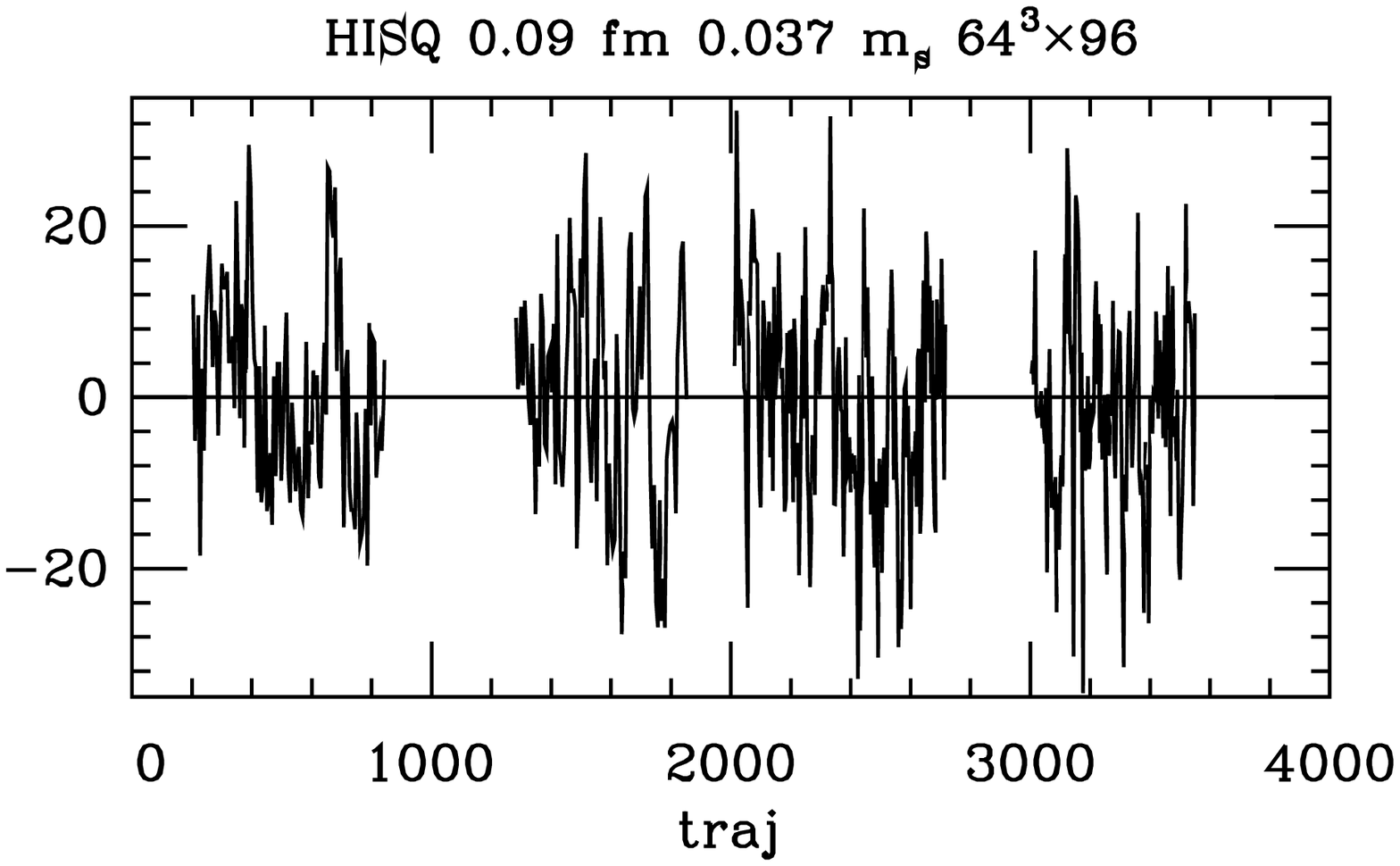}}}
&
\vtop{
\vspace{-1.16in}
\hbox{
\includegraphics[height=0.142\textwidth]{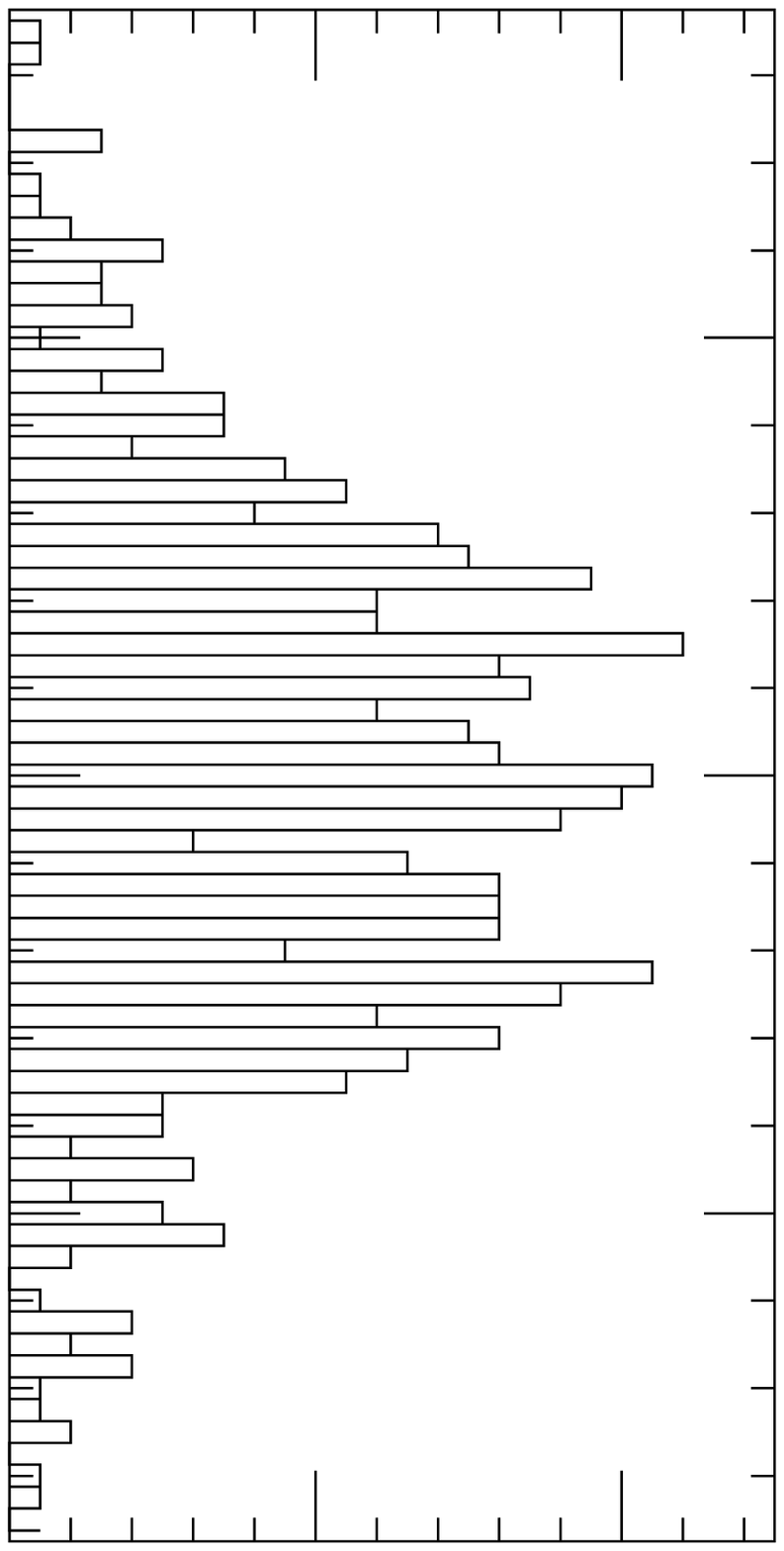}}} \\
\vtop{
\hbox{ \includegraphics[height=.20\textwidth]{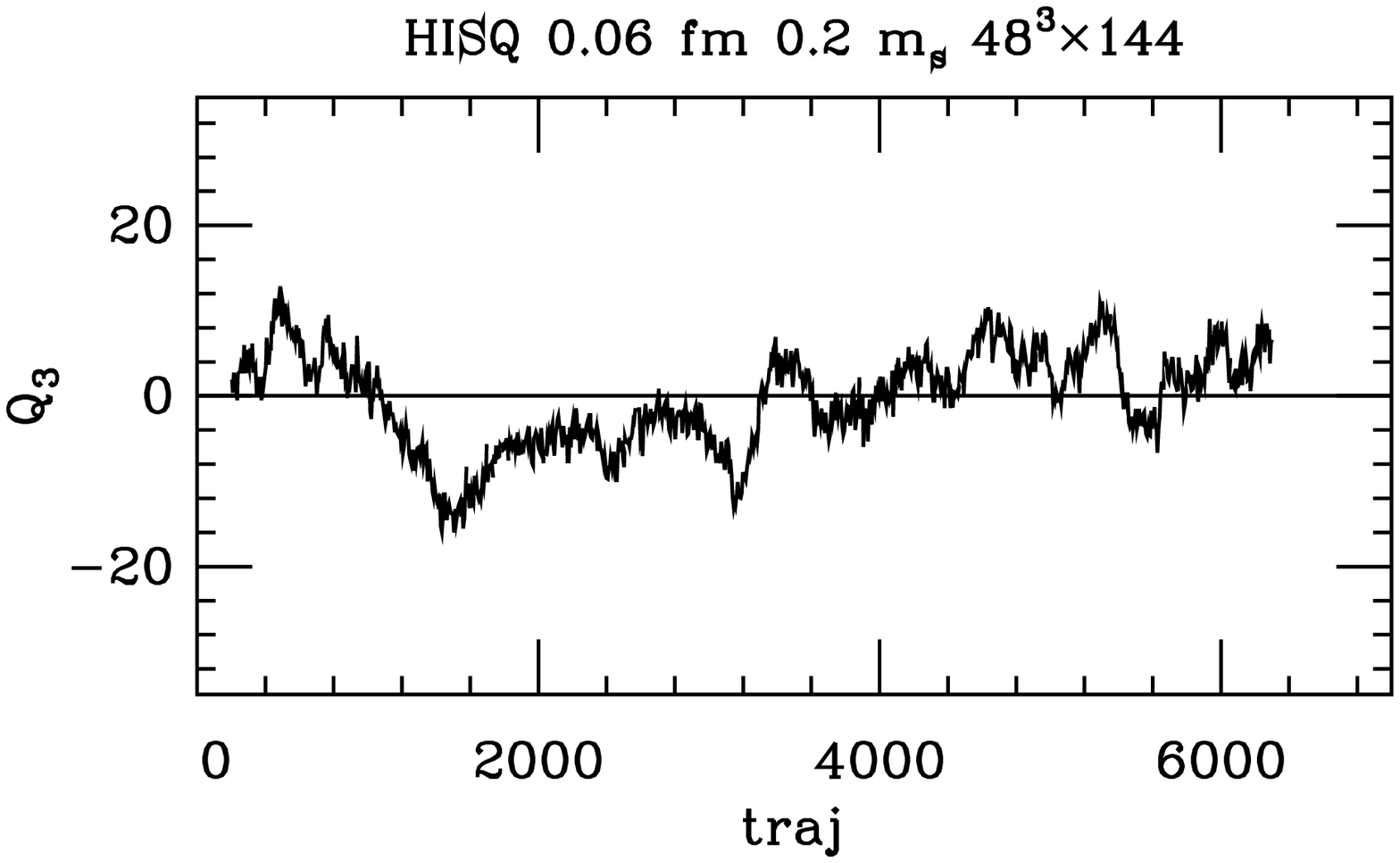}}}
&
\vtop{
\vspace{-1.16in}
\hbox{
\includegraphics[height=0.142\textwidth]{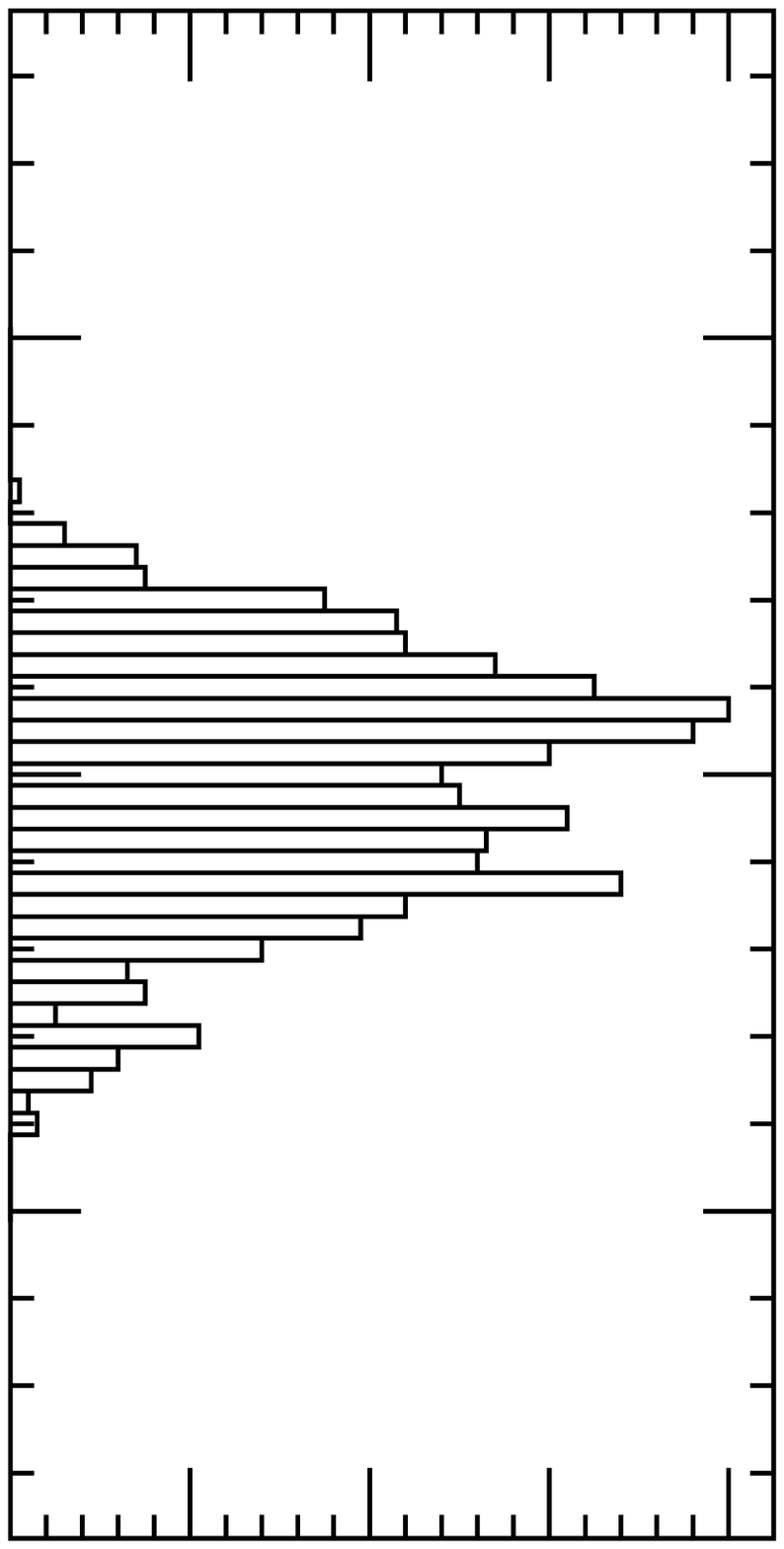}}}
&   &   \\
\end{tabular}
\caption{The time history of the topological charge for five HISQ
gauge configuration ensembles. The histogram to the right of each
time history shows the distribution of charges.
\label{fig:topo_hist}
}
\end{figure}

As in Ref.~\cite{ref:topo_susc2}, we reduce the variance in the integral
in Eq.~(\ref{eq:corrint}) by fitting $C(r)$ at large $r$ to a model that
includes the required contributions from the $\eta$ and $\eta^\prime$
mesons.  The relative strength of the two contributions is also fixed
using tree-level chiral perturbative mixing.  The integral is then
evaluated as
\begin{equation}
  \chi_t = \int_{r<r_{\rm cut}} d^4x \, C(r) + \int_{r > r_{\rm cut}} d^4x \, C(r) \, ,
\end{equation}
where the contribution for $r > r_{\rm cut}$ is derived from the
asymptotic fit and for $r < r_{\rm cut}$ it is based on the raw data.
In practice, we take $r_{\rm cut} \approx 1.2$ fm.

An interesting question for any lattice QCD simulation is whether
the tunneling rate between topological charge sectors is sufficient
to bring those sectors to equilibrium. To investigate this question,
we show in Fig.~\ref{fig:topo_hist} the time history of the topological
charge for five HISQ ensembles: those with $m_l=m_s/5$ and $a\approx 0.12$,
0.09 and 0.06~fm, and with $m_l=m_s/27\approx 0.037\, m_s$ 
for $a\approx 0.12$ and 0.09~fm.
(We do not yet have sufficient data to make a similar plot for the 
$a\approx 0.06$~fm, physical quark-mass ensemble). To the right of each
time history plot, we show a histogram of the topological charge. As expected,
the tunneling rate decreases with the lattice spacing, whereas the
fluctuations of the charge increase as the light quark mass is decreased.
It is clear from this figure that the simulation
is exploring a wide range of topological charges.
The topological susceptibility determined from the widths of these histograms
is consistent with the results in Table~\protect\ref{tab:HISQtopo}, 
although with larger statistical errors.

\begin{figure}
\centerline{\includegraphics[width=0.7\textwidth]{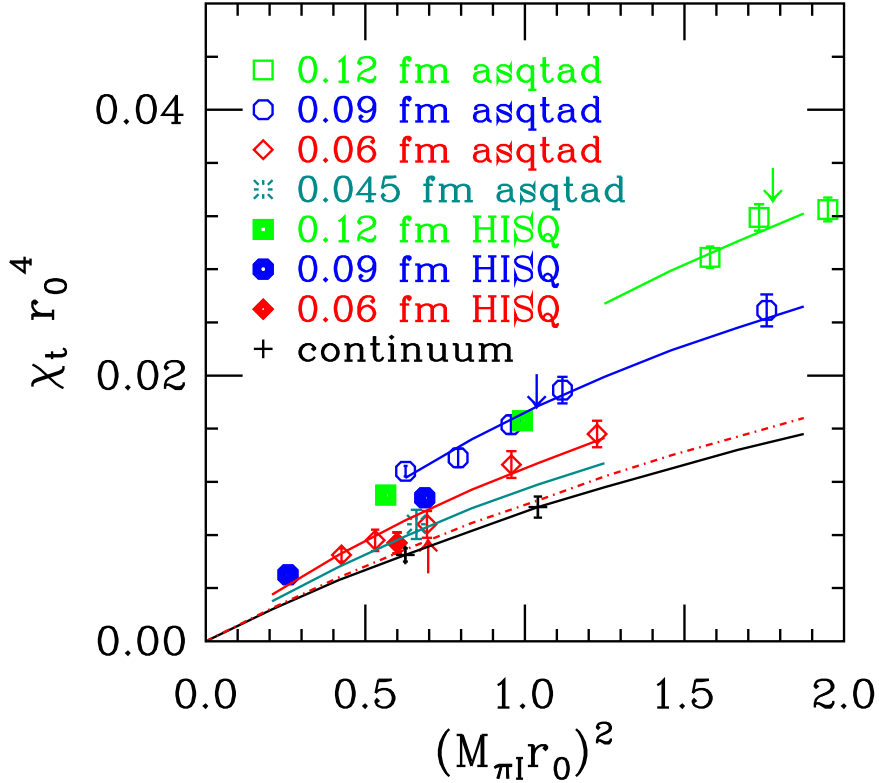}}
 \caption{Topological susceptibility {\it vs.} taste-singlet pion
   mass squared in units of $r_0$, 
   comparing results from five HISQ ensembles (filled symbols) with previously
   published asqtad results~\protect\cite{ref:topo_susc2} (open symbols). 
   Solid curves are from a joint chiral/continuum fit to the asqtad data for 
   the four lattice
   spacings shown in the figure.  The lowest (black) curve
   indicates the resulting continuum extrapolation of the asqtad
   fit with two representative points displaying the
   extrapolation errors.  The (red) dot-dashed curve shows the leading
   order prediction in chiral perturbation theory.  The (green) arrow
   above the (green) 0.12~fm asqtad curve indicates the asqtad point
   with a light quark mass comparable to that of the upper solid (green) 
   HISQ square.  Similarly, 
   the (blue) arrow above the (blue) 0.09~fm asqtad curve indicates the
   asqtad point with a light quark mass comparable to the upper solid (blue)
   HISQ octagon, and the (red) arrow below the (red) 0.06~fm asqtad curve 
   locates the asqtad point with a light quark mass comparable to the 
   solid (red) HISQ diamond.
\label{fig:HISQtopo}
}
\end{figure}

Results are tabulated in Table~\ref{tab:HISQtopo} and compared with those
for asqtad in Fig.~\ref{fig:HISQtopo}. The comparison with the asqtad
results provides a clear demonstration of the improvement in the HISQ
configurations. Because the susceptibility is computed without involving
valence quarks, this comparison directly tests whether the change
in sea quark action leads to the expected improvement in the gauge
configurations. We observe in Fig.~\ref{fig:HISQtopo} that
the HISQ points with $a\approx 0.12$, 0.09 and 0.06~fm are near
the asqtad curves with $a\approx 0.09$, 0.06 and 0.045~fm, respectively.
The HISQ points are to the left of the corresponding asqtad points
because the horizontal axis
is the mass of the taste singlet pion (the heaviest pion taste), and the
reduction in taste symmetry breaking moves the HISQ points to the left. It is
the decrease of the susceptibility for the HISQ configurations relative to
those of the asqtad configurations that represents the improvement in the
gauge configurations.

\section{Taste symmetry}
\label{TASTE}

\begin{figure}[t]
\includegraphics[width=0.6\textwidth]{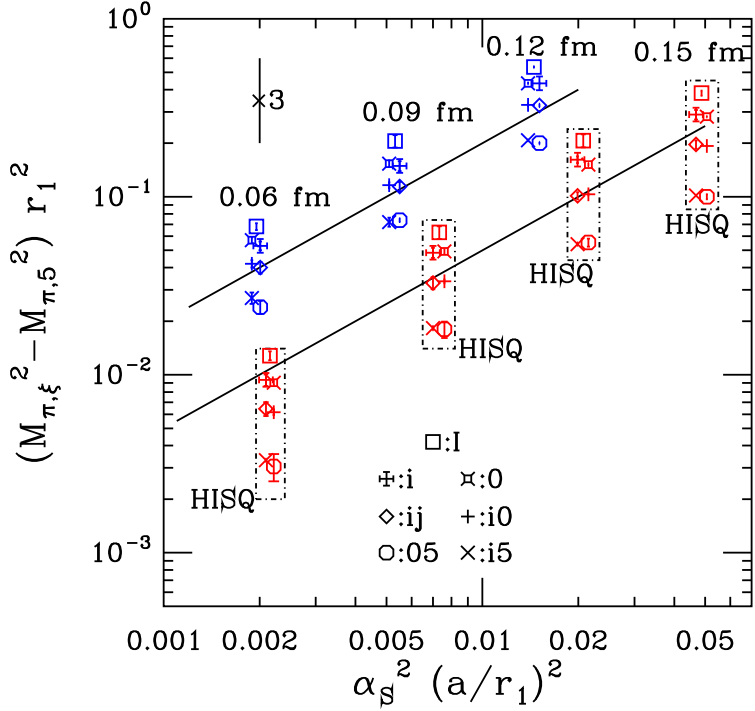}
 \caption{Pion taste splitting of pions for asqtad (blue) and HISQ (red) 
actions.  For clarity, the HISQ splittings are also enclosed in dashed-dotted 
boxes, and nearly degenerate masses have been displaced slightly
in the horizontal direction.
Differences between the squared masses of various taste pions and that of the
Goldstone  pion are shown in units of $r_1$, and plotted versus
the expected leading dependence of taste violations in the theory, 
$\alpha_S^2 a^2$, also in $r_1$ units. Here, we use $\alpha_S=\alpha_V$ at scale $q^*\!=\!2/a$.
The two diagonal lines are not fits, but merely lines with slope 1, showing the expectation
if the splittings are linear in
$\alpha_S^2 a^2$.  The vertical line at the upper left shows the displacement
associated with a factor of three in splittings. The numerical values of
the HISQ taste splittings plotted here are given in 
Table~\ref{tab:taste_degenerate} of the appendix. 
\label{fig:taste-split-2oa}
}
\end{figure}

For the $m_l=m_s/5$ ensembles
at each lattice spacing, we have measured the masses of pseudoscalar
mesons of all 16 possible tastes:  $I$, $\xi_\mu$, $\xi_{\mu\nu}$,
$\xi_{\mu5}$, and $\xi_5$, with $\xi_{\mu\nu}\equiv [\xi_\mu,\xi_\nu]/2$
and $\xi_{\mu5}\equiv \xi_\mu\xi_5$.
  Figure \ref{fig:taste-split-2oa}
shows the difference in squared mass between pions of a given taste
and that of the Goldstone pion (taste $\xi_5$), plotted
versus $\alpha_S^2a^2$, the expected dependence of the leading
taste-violating effects.  
The splittings for the corresponding
asqtad ensembles are also shown for comparison.  
Here ``pion'' means the
unitary meson with degenerate valence masses equal to $m_l$.  Tastes
are labeled by their index or indices; \ie\ $ij$ denotes taste $\xi_{ij}$,
with $i$ and $j$ labeling generic spatial directions.  Results for
different spatial taste indices have been averaged since exact (discrete)
lattice symmetries imply that their masses are equal. In most cases, the 
differences between
squared masses have been calculated with a jackknife procedure.

One can see that the HISQ splittings are typically a factor of three smaller 
than the asqtad ones at the same lattice spacing. In fact the improvement
with HISQ increases as the lattice spacing decreases, so that at $a\approx0.06$
fm, the HISQ splittings are a factor of 6 to 8 smaller than the asqtad ones.
Indeed the HISQ splittings at a given lattice spacing are comparable
to, but somewhat
smaller than, the asqtad splittings at the next smaller lattice spacing, \ie\
HISQ splitting at $a\approx 0.09$ fm are a bit smaller than asqtad splittings 
at $a\approx 0.06$ fm.

\newcommand{\R}[1]{{\color{red}#1}}
\newcommand{\G}[1]{{\color{green}#1}}
\newcommand{\B}[1]{{\color{blue}#1}}

For the strong coupling $\alpha_S$, 
we use $\alpha_V$,
the coupling from the asqtad heavy-quark
potential, calculated in Ref.~\cite{Davies:2002mv}
at next-to-leading order in tadpole-improved lattice perturbation theory
\cite{Lepage:1992xa}:
\begin{equation}\eqn{alphaV}
-\ln({\rm plaq}) = 3.0682(2)\,\alpha_V (3.33/a)\;[1+\alpha_V\{-0.770(4) - 0.09681(9)\;n_f \}]\ ,
\end{equation}
where plaq is the average plaquette of the ensemble, $n_f$ is the number of sea-quark flavors,
and the scale $q^*=3.33/a$ for the plaquette
is set by the BLM procedure \cite{Brodsky:1982gc}.  For taste-violations with asqtad quarks, one expects
the appropriate scale to be somewhat lower, since the asqtad smearings are designed to remove coupling
of the quarks to gluons with any momentum component equal to $\pi/a$. In
Fig.~\ref{fig:taste-split-2oa}, we choose scale $q^*=2/a$ for $\alpha_V$ (the adjustment of scale is
made using the universal two-loop formula).
For the HISQ action, the perturbative calculation corresponding to \eq{alphaV}
has not to our knowledge been performed.  However, since the dependence
on $n_f$ in \eq{alphaV} is fairly small, we think it is reasonable
also to use the asqtad formula for HISQ, at least in this qualitative comparison
of discretization effects.  One can see that the HISQ points are shifted
to the right relative to corresponding asqtad ones; this is the effect of
using $n_f=4$ rather than $n_f=3$.

By comparing the asqtad splittings 
in Fig.~\ref{fig:taste-split-2oa} with the upper line (which has slope 1),
one can see that the asqtad splittings are almost
proportional to $\alpha_V^2(2/a)\;a^2$, but fall very slightly faster as $a$ 
decreases.  The HISQ splittings fall still more rapidly at the smallest lattice
spacings, presumably because the proper choice of $q^*$ is significantly
smaller in the HISQ case.  That is reasonable, since the greater smearing
present in the HISQ action should produce greater damping of the couplings
of gluons to quarks at high momenta. On
the other hand,  the HISQ splittings fall more slowly than $\alpha_V^2(2/a)\;a^2$
at the coarsest lattice spacings (between 0.15 fm and 0.12 fm), which is evidence for 
higher order ($\alpha_S^3a^2$ or $a^4$) contributions.

\begin{figure}
\includegraphics[width=0.6\textwidth]{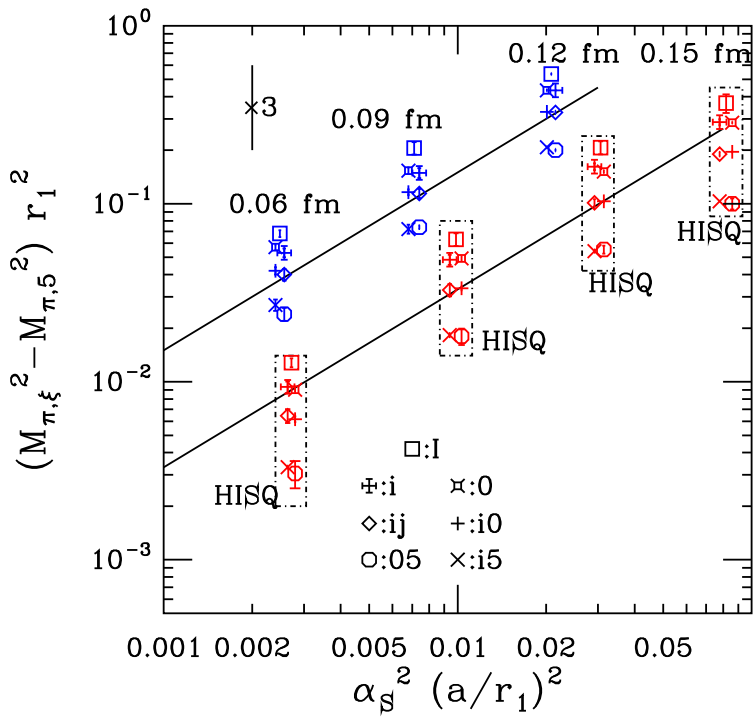}
 \caption{Same as Fig.~\protect\ref{fig:taste-split-2oa}, but with $\alpha_S=
\alpha_V(q^*\!=\!1.5/a)$.  
\label{fig:taste-split-1.5oa}
}
\end{figure}

For comparison, Fig.~\ref{fig:taste-split-1.5oa} shows the effect of choosing
a somewhat lower scale, $q^*=1.5/a$, in other words $\alpha_S=\alpha_V(1.5/a)$. 
Now the asqtad splittings
fall slightly slower than $\alpha_S^2a^2$ as $a$ decreases, while the
HISQ splittings are closer to linear, but still fall faster between
0.09 fm and 0.06 fm.  We could continue to reduce $q^*$ until this drop
became linear, but it would be an arbitrary exercise since we do not have
the correct perturbative  formula in the HISQ case.

In \figrefs{taste-split-2oa}{taste-split-1.5oa}, both the HISQ and asqtad
masses show an approximate $SO(4)$ taste symmetry: The masses form
five multiplets with tastes P, A, T, V, and I (pseudoscalar, axial-vector,
tensor, vector and singlet tastes).  
This is an ``accidental'' symmetry,
because the exact lattice
symmetries do not require this structure, but would allow all eight multiplets
listed in the legends to be non-degenerate.  The origin
of the $SO(4)$ taste symmetry of pions \cite{LEE_SHARPE}
is explained briefly below.

\begin{figure}
\includegraphics[width=0.6\textwidth]{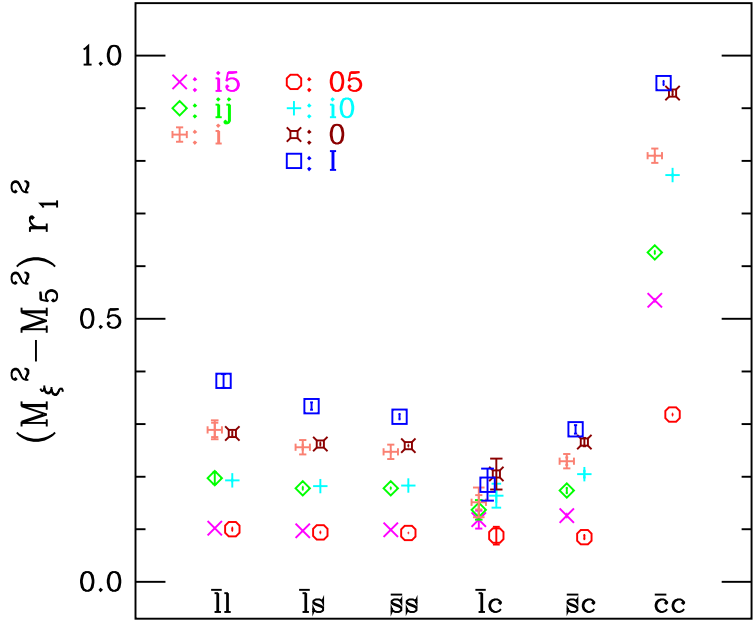}
 \caption{Meson taste splittings
on the $a\approx0.15\;$fm,  $m_l=m_s/5$ ensemble. 
As in Fig.~\protect\ref{fig:taste-split-2oa}, the squared mass splitting 
between pseudoscalar mesons of different tastes 
and the lightest one with taste $\xi_5$ (the Goldstone pion
for the $\bar{l} l$ case) is given in units of $r_1$.  The types of quarks 
in the mesons are shown on the abscissa: $l$, $s$, and $c$ stand for
light ($u,d$), strange, and charm quarks, respectively. All mesons
are the unitary ones, with each valence quark mass equal to one of the sea 
quark masses.  Note, however, that all mesons here are treated as 
flavor-charged, so that
even in the  $\bar ss$ and $\bar cc$ cases, no quark-disconnected diagrams 
are included. The numerical values of most of these taste splittings are given
in Tables~\ref{tab:taste_degenerate} and \ref{tab:taste_heavylight} of the
appendix.  
\label{fig:all-mesons-a15}
}
\end{figure}

The $SO(4)$ symmetry is seen even more clearly in 
the $\bar ll$ (pion) masses in \figref{all-mesons-a15}. Up to quite small errors, no breaking
of the symmetry is visible. 
The $\bar ll$ masses also obey a rough ``equal spacing rule'' of squared
masses  between
the P, A, T, V, and I tastes (and with that ordering).  The equal spacing is familiar
from the asqtad case \cite{RMP}. It arises in staggered chiral perturbation 
theory (\schpt)
\cite{LEE_SHARPE,SCHPT} from the fact that the dominant taste-breaking
chiral operator (\ie\ the one with the largest coefficient) is
\begin{equation}
\label{eq:O4}
\cO_4 = a^2 \Tr(\xi_{\lambda5}\,\Sigma\;
        \xi_{5\lambda}\,\Sigma) + h.c.
\end{equation}
where $h.c.$ stands for the Hermitian conjugate, and 
$\Sigma=\exp(i\Phi/f)$, with $\Phi$ the pion field and $f$ the LO pion decay constant.

Operators such as $\cO_4$
are representatives, at the chiral level,  of taste-symmetry-breaking 
four-quark operators 
in the Symanzik effective
theory. The four-quark operators may be labeled by the 
spin and taste of the bilinears from which they are constructed. 
In particular, $\cO_4$ is generated by the 
operators $[I\times A]$, $[T\times A]$ and $[P\times A]$, where, for example
\begin{equation}
\label{eq:TxA}
[T\times A]  \equiv a^2\, (\bar q\,\gamma_{\mu\nu}\otimes \xi_{\lambda5}\, q)\;
(\bar q\,\gamma_{\nu\mu}\otimes \xi_{5\lambda}\, q)\ ,
\end{equation}
with $\gamma_{\mu\nu}\equiv [\gamma_\mu,\gamma_\nu]/2$.
This notation is basically that of Lee and Sharpe \cite{LEE_SHARPE}, except that
we use
$I$ instead of $S$ to denote scalar spin or taste, to avoid confusion with $s$
for ``strange.''  
The four-quark operators that give rise to $\cO_4$ are what are known
as ``type-A'' operators, which are invariant under $SO(4)$ of taste, as well
as the $SO(4)$ of Euclidean space-time rotations. There are also ``type-B'' operators
that couple spin and taste indices, and break these $SO(4)$ symmetries down to
a diagonal subgroup of joint spin-taste $90^\circ$ rotations. An example is
\begin{equation}
\label{eq:TmuxAmu}
[T_\mu\times A_\mu]  \equiv a^2 \sum_{\mu=1}^4\;
(\bar q\,\gamma_{\mu\nu}\otimes \xi_{\mu5}\, q)\;
(\bar q\,\gamma_{\nu\mu}\otimes \xi_{5\mu}\, q)\ .
\end{equation}
The accidental $SO(4)$ taste symmetry of the pions results from the fact that
the type-B operators do not have chiral representatives at LO. 
The non-trivial  space-time structure in the type-B case requires more 
than two derivatives in
the chiral operators, making their representatives next-to-leading order (NLO)
in the chiral expansion \cite{LEE_SHARPE}.

Moving up in meson mass in \figref{all-mesons-a15}, one sees that $SO(4)$ 
is still
a good symmetry for the $\bar ls$ (kaon) and $\bar ss$ states, although some 
small symmetry violations are coming into view in the $\bar ss$ case.  
The $SO(4)$ violations for $\bar ss$ can be seen most clearly in
Table~\ref{tab:taste_degenerate}, where they are significant (though still
quite small) at $a\approx 0.15$~fm. 
This is reasonable, since, as both the mass and lattice spacing increase,
NLO chiral corrections are less suppressed,
and these can violate $SO(4)$.

At still higher mass, ordinary light-meson \schpt\ ceases to be applicable,
and a heavy-light version is necessary.  For the case here, where both
light and heavy quarks have the staggered action, the
heavy-light chiral theory has recently been worked out \cite{Komijani:2012fq}.
In that theory (``all-staggered heavy meson chiral perturbation theory'' -- \aschpt), 
the LO heavy-light chiral Lagrangian 
is of order $k$, the heavy-light meson residual
momentum, which is taken to be of the order 
of the pion momentum $p\sim M_\pi$. This is different from the light-light
case, in which the LO Lagrangian is order $M_\pi^2$. 
Taste violations are LO in the light-light case because the taste splittings 
in squared masses can be comparable to $M_\pi^2$; in other words we assume
$a^2\sim M_\pi^2$  (with appropriate factors of $\Lambda_{QCD}$ inserted to
make the dimensions the same).  In the heavy-light case, on the other
hand, taste violations are NLO since $a^2\ll M_\pi$.
This rough picture is actually 
consistent with what is seen in \figref{all-mesons-a15}, 
where the splittings in squared masses remain comparable from the 
$\bar ll$ case through the $\bar sc$ case, and in 
Tab.~\ref{tab:taste_heavylight}, where we show the splittings in the
$\bar l c$ and $\bar s c$ systems for the $a\approx 0.15$~fm, $m_l=m_s/5$
ensemble.  The splittings in the masses 
themselves are thus much smaller for $\bar lc$ or $\bar sc$ mesons than they 
are for $\bar ll$ mesons.
For example, the measured 
taste splitting between the root-mean-squared (RMS) $D_s$ meson and the lightest
(taste $\xi_5$) $D_s$ meson  at $a\approx0.12\;$fm is only about 10~MeV, 
while it is about 110~MeV when the taste $\xi_5$ pion takes its physical mass.
Fortunately, it is possible to show in \aschpt\ that the
one-loop diagrams give taste-invariant masses to the heavy-light mesons, 
even though the diagrams contain pion propagators that break taste symmetry.  
This means that
all taste-violations in the heavy-light masses at NLO come from analytic terms
in the \aschpt\ Lagrangian, and may be analyzed straightforwardly.

There are other key differences between the light-light and heavy-light chiral
theories.
The non-relativistic nature of the heavy quark in heavy-light systems
breaks space-time $SO(4)$ invariance and
thereby introduces the four-velocity $v^\mu$ of the heavy quark explicitly
into the heavy-light chiral theory. Factors of $v^\mu$ can 
substitute for derivatives
in the chiral Lagrangian, and thereby allow type-B operators to have 
chiral representatives that are the same order in the chiral expansion 
as those from type-A operators. In addition, heavy-quark spin symmetry, 
which produces (approximate) degeneracy of the pseudoscalar (\eg\ $D$) and vector (\eg\ $D^*$) mesons,
introduces spin degrees of freedom into the chiral theory. 
The gamma matrix
$\gamma_\mu$ can then play a role similar to that of $v^\mu$, also
allowing type-B chiral operators to appear at the same order as 
type-A operators.  The presence of type-B operators is visible in the
breaking of $SO(4)$ symmetry for the heavy-light
$\bar lc$ and $\bar sc$ mesons in \figref{all-mesons-a15}. The splittings 
within $SO(4)$
multiplets are particularly clear in the $\bar sc$ case, where the 
statistical errors are smaller.

One can go further and study the particular pattern of taste splittings (both
$SO(4)$ invariant and $SO(4)$ breaking) for heavy-light mesons.
For this discussion,
we assume that the lattice is sufficiently fine, or the charmed quark is sufficiently
improved, that it may be treated as a ``continuum-like,'' and corrections
of order $(am_c)^2$ may be neglected. This means that the contributions of the
heavy quark to the Symanzik effective theory are identical to those of a light quark. In particular,
the same four-quark operators that dominated for light quarks, namely
$[I\times A]$, $[T\times A]$ and $[P\times A]$, will be the dominant type-A
operators in the heavy-light case. Taste splittings of heavy-light meson masses
can come only from the ``heavy-light'' versions of these operators,
$[I\times A]_{hl}$, $[T\times A]_{hl}$ and $[P\times A]_{hl}$,
which  couple heavy and light bilinears, \eg\
\begin{equation}
\label{eq:TxAhl}
[T\times A]_{hl}  \equiv a^2\, (\bar Q\,\gamma_{\mu\nu}\otimes \xi_{\lambda5}\, Q)\;
(\bar q\,\gamma_{\nu\mu}\otimes \xi_{5\lambda}\, q)\ ,
\end{equation}
where $Q$ and $q$ are the heavy- and light-quark fields, respectively.

In \aschpt, we define the heavy-light meson 
field $H_j$ by%
\footnote{For ease of comparison with the light-light case, we discuss the 
heavy-light 
chiral theory in Euclidean space, in contrast to what was done in Ref.~\protect\cite{Komijani:2012fq}.}
\begin{equation} \label{eq:H_definition}
  H_{j} = \frac{1 + \vslash}{2}\left[ \gamma_\mu D^{*}_{\mu j}
    + i \gamma_5 D_{j}\right]\ ,
\end{equation}
where $j$ labels the light flavor, $D_j$ and $D^{*}_{\mu j}$ are fields that
annihilate pseudoscalar and vector mesons, respectively, and
$H_{j}$, $D_j$ and $D^{*}_{\mu j}$ are all $4\times4$ matrices in taste space.
Then the four-quark operators above have the following chiral representatives:
\begin{eqnarray}
\big[I\times A\big]_{hl} & \rightarrow & a^2 
                   \Tr\left(\overline{H} \xi_{\lambda5} H \xi_{5\lambda}  \right)\ , \nonumber \\
\big[P\times A\big]_{hl} & \rightarrow & 0\ , \eqn{HLtypeA} \\
\big[T\times A\big]_{hl} & \rightarrow & a^2 
                   \Tr\left(\overline{H} \gamma_{\mu\nu} \xi_{\lambda5} 
                   H \gamma_{\nu\mu} \xi_{5\lambda} \right)\ , \nonumber 
\end{eqnarray}
where $\Tr$ is the trace over Dirac, taste, and flavor indices.  Here we have set
the pion field $\Phi$ to zero since we are not interested in couplings to pions, and have omitted
two-trace terms, which vanish in that limit. The operator
$[P\times A\big]_{hl}$ has no chiral
representative in the heavy-light case because 
$(1+\vslash)\gamma_5(1+\vslash)=0$.
(Only ``upper components'' of the heavy-quark field appear, while $\gamma_5$ couples upper and
lower components.)
Performing the traces in \eq{HLtypeA}, the chiral operators give the relative
contributions to the heavy-light pseudoscalar masses shown 
in~\tabref{TasteSplittingA}. (The contributions from the two operators are
proportional to each other because $\gamma_{\mu\nu}$ in \eq{HLtypeA}
commutes with $\gamma_5$ from \eq{H_definition}.) 
Note that the same equal-spacing rule and P, A, T, V, I 
ordering that appeared for light-light mesons
also appears in the heavy-light case. This overall pattern is evident in the heavy-light
lattice data in \figref{all-mesons-a15}, although of course the taste-$SO(4)$ 
breaking produces additional structure.

\begin{figure}
\includegraphics[width=0.6\textwidth]{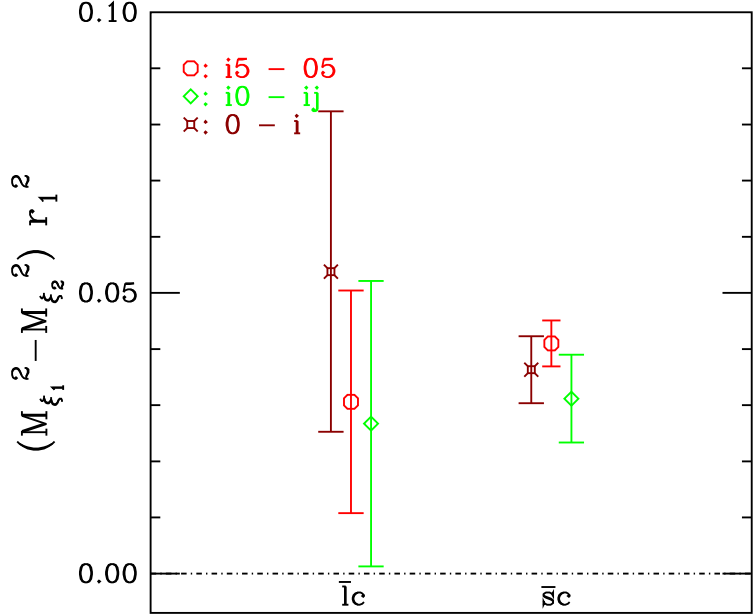}
 \caption{Heavy-light taste splittings that break $SO(4)$ taste symmetry
on the $a\approx0.15\;$fm,  $m_l=m_s/5$ ensemble. 
As in
 Fig.~\protect\ref{fig:all-mesons-a15}, $\bar lc$ is a light-charm meson, and
$\bar sc$ is a strange-charm meson.  
The errors in the squared-mass differences
have been calculated with a jackknife procedure. The numerical values
of these splittings are given in Table~\ref{tab:taste_heavylight}. 
\label{fig:SO4-breaking}
}
\end{figure} 
\begin{table}
\caption{Relative contributions to the masses of heavy-light 
mesons of given tastes due to the (apparently) dominant type-A operators, \protect\eq{HLtypeA}.}
\label{tab:TasteSplittingA}
\setlength{\tabcolsep}{9pt}
\begin{center}
\begin{tabular}{|c|c|c|c|c|} \hline
$\xi_5$ & $\xi_{5\mu}$  & $\xi_{\mu\nu}$ 
& $\xi_\mu$ &  $I$ \\ \hline 
-4  &  -2 &  0 &  +2 &  +4 \\ \hline 
\end{tabular}
\end{center}
\end{table}

For type-B operators, we have no experience from the light-light case as to which one or
ones might give the largest contributions. However, we can guess that the 
four-quark operator $[T_\mu\times A_\mu\big]_{hl}$ would be dominant, since
it is the only type-B operator that has the same spin and taste as one of the dominant
type-A operators. The chiral operator generated is
\begin{equation}
\eqn{HLtypeB} 
\big[T_\mu\times A_\mu\big]_{hl}  \rightarrow  a^2 \sum_\mu
\Tr\left(\overline{H} \gamma_{\nu\mu} \xi_{5\mu} H \gamma_{\mu\nu} \xi_{\mu5} \right) \ , 
\end{equation}
where the same simplifications as in \eq{HLtypeA} have been made. 
\tabref{TasteSplittingB} shows the pattern of mass splittings resulting from
this operator.  We have assumed that the overall sign of the operator is the same
as that of the (net) type-A operator that generated \tabref{TasteSplittingA}.  
Note first that the overall effect on the
``centers of gravity'' of the $SO(4)$
multiplets (average masses, taking into account multiplicities) is the same as for the
type-A operators in \tabref{TasteSplittingA}. 

\begin{table}
\caption{Relative contributions to the masses of heavy-light mesons 
of given tastes due to the (apparently) 
dominant type-B operator, \protect\eq{HLtypeB}.}
\label{tab:TasteSplittingB}
\setlength{\tabcolsep}{9pt}
\begin{center}
 \begin{tabular}{|c|c c|c c|c c|c|} \hline
$\xi_5$ & $\xi_{05}$ & $\xi_{i5}$ & $\xi_{ij}$ 
& $\xi_{i0}$ & $\xi_i$ & $\xi_0$ & $I$ \\ \hline 
-6 &-6 &-2 &-2 &+2 &+2 &+6 & +6\\ \hline 
\end{tabular}
\end{center}
\end{table}

The main implication of 
\tabref{TasteSplittingB}, however, is the pattern of $SO(4)$ violations it predicts. 
In the axial taste multiplet, the 
spatial
component $\xi_{i5}$ is raised relative to the time component $\xi_{05}$, but this
situation is reversed in the tensor and vector multiplets.  Furthermore, the absolute value
of the time-space taste splitting is the same in each of the three multiplets.  This structure
is exactly what is observed in \figref{all-mesons-a15}.  We make the comparison
with the lattice data more quantitative
in \figref{SO4-breaking}, where  the $SO(4)$-breaking mass differences are plotted for the
$\bar lc$ and $\bar sc$ cases.  These mass differences have been calculated directly
with a jackknife procedure (rather than simply propagating the errors from the differences
with the Goldstone meson shown in~\figref{all-mesons-a15})
in order to take advantage of the correlations to reduce the error in the splittings.  
In the $\bar sc$ case, there is
good evidence that the splittings have the sign and relative magnitude predicted 
in \tabref{TasteSplittingB}.  The $\bar lc$ case is much noisier, but the signs and magnitudes
are at least consistent with expectations. Note also that there is no evidence for a
light-mass dependence of the splittings, which is as expected at this order in \aschpt.

There is no chiral effective theory to analyze the splittings for heavy-heavy mesons (the $\bar cc$ case in
\figref{all-mesons-a15}).
Nevertheless, it is interesting to see that the $SO(4)$ breaking,
already clear in the $\bar sc$ and $\bar lc$ cases,
gets very strong in the $\bar cc$ case.  In particular, the spacings
between some members of different
$SO(4)$ multiplets ($\xi_0$ and $I$, or $\xi_{i0}$ and $\xi_i$), are 
smaller than the splittings within multiplets.

\section{Thermalization and Autocorrelation times}
\label{AUTO}

\begin{figure}[t]
\centerline{\includegraphics[width=0.5\textwidth]{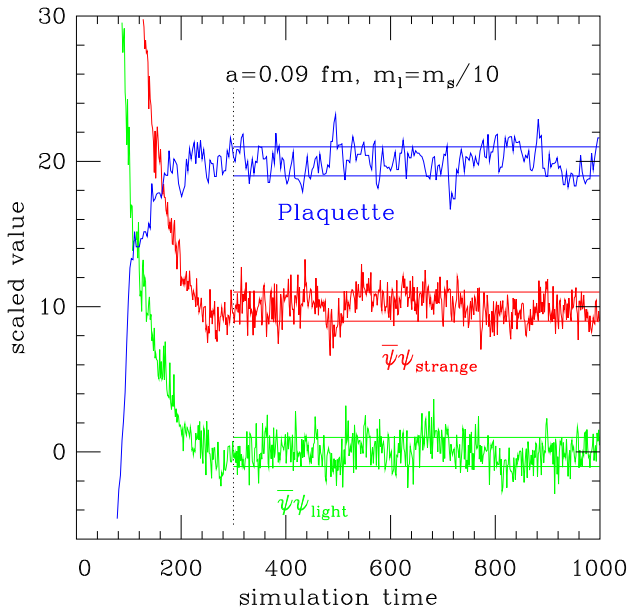}}
\caption{
Equilibration of the plaquette and $\bar\psi\psi$ in the ensemble with
$a\approx 0.09$~fm and $m_l=m_s/10$.  The quantities are rescaled so that 
their standard deviations, obtained from the equilibrated
part of the run, are one, and shifted so that the distributions center
at zero, ten and twenty.  Horizontal lines for each quantity show
the $\pm$ one standard deviation range of the equilibrated quantity,
obtained from averaging over simulation times 300 to 6300. (Only the
first 1000 time units are shown here.)  The vertical line at $T=300$
shows where we began taking measurements in analysis projects.
This particular ensemble was started from a configuration very far
from equilibrium, so the warm up effects are dramatic in this plot.
\label{fig:warmuphistory}
}
\end{figure}

At the start of a new simulation, the system will not be
in equilibrium, and gauge configurations generated before 
it has thermalized should not be included in measurements of
physical quantities. The amount of running discarded for 
thermalization was determined
from time histories of the plaquette and chiral condensate, and,
when available later, spectrum measurements.  The equilibration
time varies among the ensembles, and is typically 200 time units,
or 300 time units for ensembles starting far from equilibrium.
Figure~\ref{fig:warmuphistory} shows the approach to equilibrium
in the ensemble with $a\approx 0.09$ fm and $m_l=m_s/10$, which
started from a configuration very far from equilibrium.

Because each successive gauge configuration is generated from the previous
one {\it via} a molecular dynamics-based evolution in ``simulation time'', 
it will resemble the previous configuration to some degree. This 
leads to correlations between the values of various observables measured on 
configurations that are nearby in simulation time. 
If these correlations are not taken into account, statistical analysis of 
lattice data will tend to underestimate the errors,
since the standard methods for estimating covariance matrices assume that 
each sample is uncorrelated. 
We have chosen to save a gauge configuration 
every five (for $a >  0.09$ fm) or six (for $a \leq 0.09$ fm)
units of simulation time, on the grounds that observables measured on 
configurations separated by less than five units will be so strongly 
correlated that the small amount of extra information that can be gleaned 
from them is not worth the effort required to extract it. 
For many observables, however, five time units is not sufficient separation 
to eliminate autocorrelations. The choice to save a configuration every 
five or six time units represents
a middle ground: two configurations five units apart differ sufficiently 
that it is worth saving them both, but are correlated enough that these 
correlations must be considered when conducting statistical analyses.

We have computed the autocorrelations between the values of various 
observables on pairs of configurations as a function of the simulation-time 
separation of the pair. We define the autocorrelation of observable 
$\mathcal{O}$ at a simulation time separation $t$ as
\begin{equation}
C(\mathcal{O};t) = \left. \frac{\LL \mathcal{O}_{i} \mathcal{O}_{j} 
\RR - \LL \mathcal{O}_{i} \RR \LL \mathcal{O}_{j} \RR}
{\LL \mathcal{O}^2 \RR - \LL \mathcal{O} \RR^2} \right|_{j=i+t}
\label{ac-def}
\end{equation}
where the expectation values in the numerator are taken over all pairs of 
configurations $i,j$ separated by $t$ units of molecular dynamics time. 
Note that this value is unity by construction for $t=0$.

This definition suffices for observables like the average plaquette, 
which has a uniquely measurable value on each configuration. 
However, we use stochastic estimators to measure the quark scalar condensates. 
Fluctuations in the measured value of the scalar condensate between
different gauge configurations are partly due to physically-meaningful 
fluctuations of the gauge fields, and partly due to fluctuations in the 
stochastic estimator. Na\"{i}ve application of Eq.~(\ref{ac-def}) to such 
quantities will underestimate the degree of autocorrelation between the 
scalar condensates
on nearby lattices, since the numerator includes extra fluctuations from the 
variation in the stochastic estimator on different configurations, while the 
denominator does not.

\begin{figure}[t]
\includegraphics[height=0.48\textwidth]{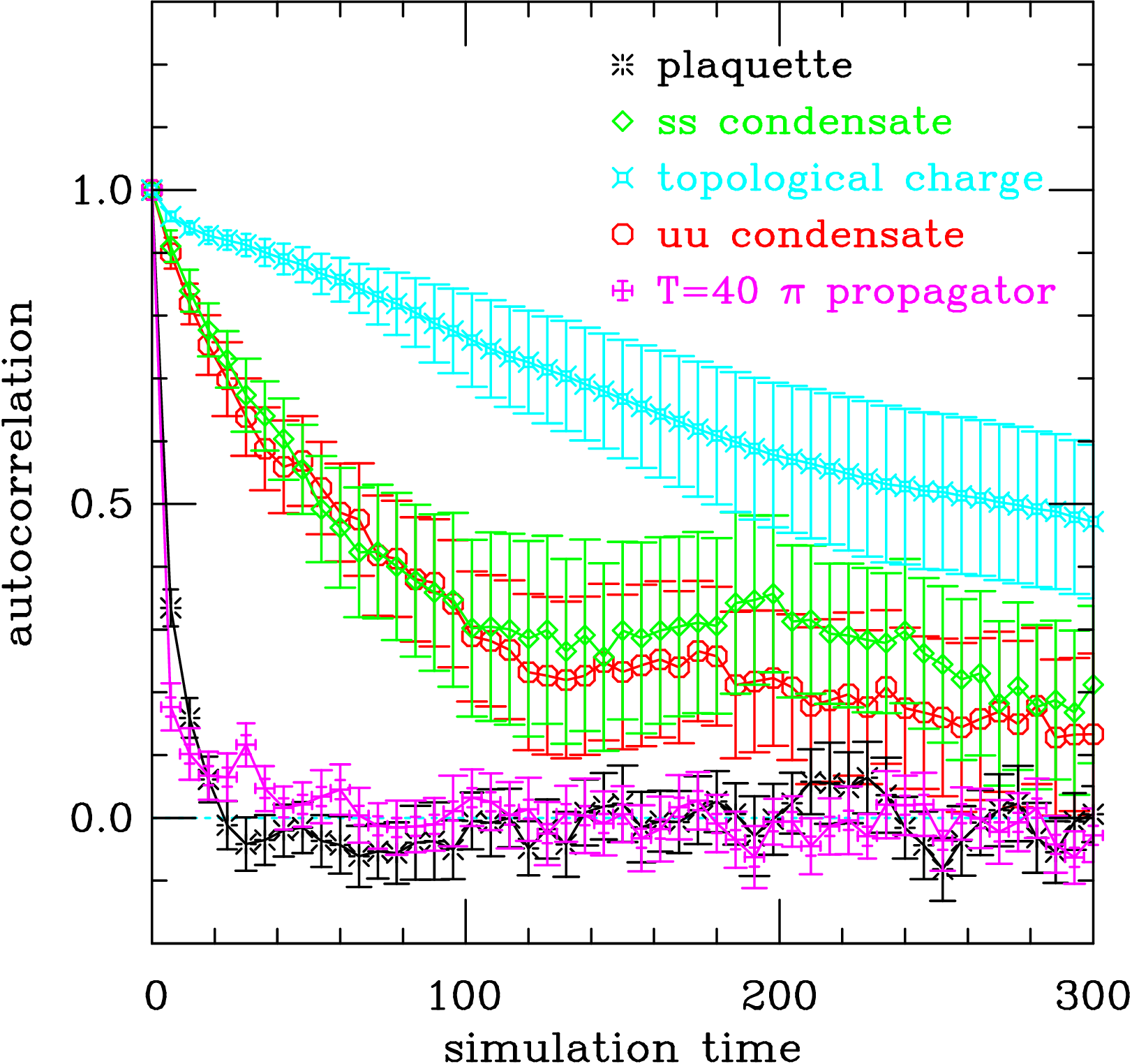}\hfil
\includegraphics[height=0.48\textwidth]{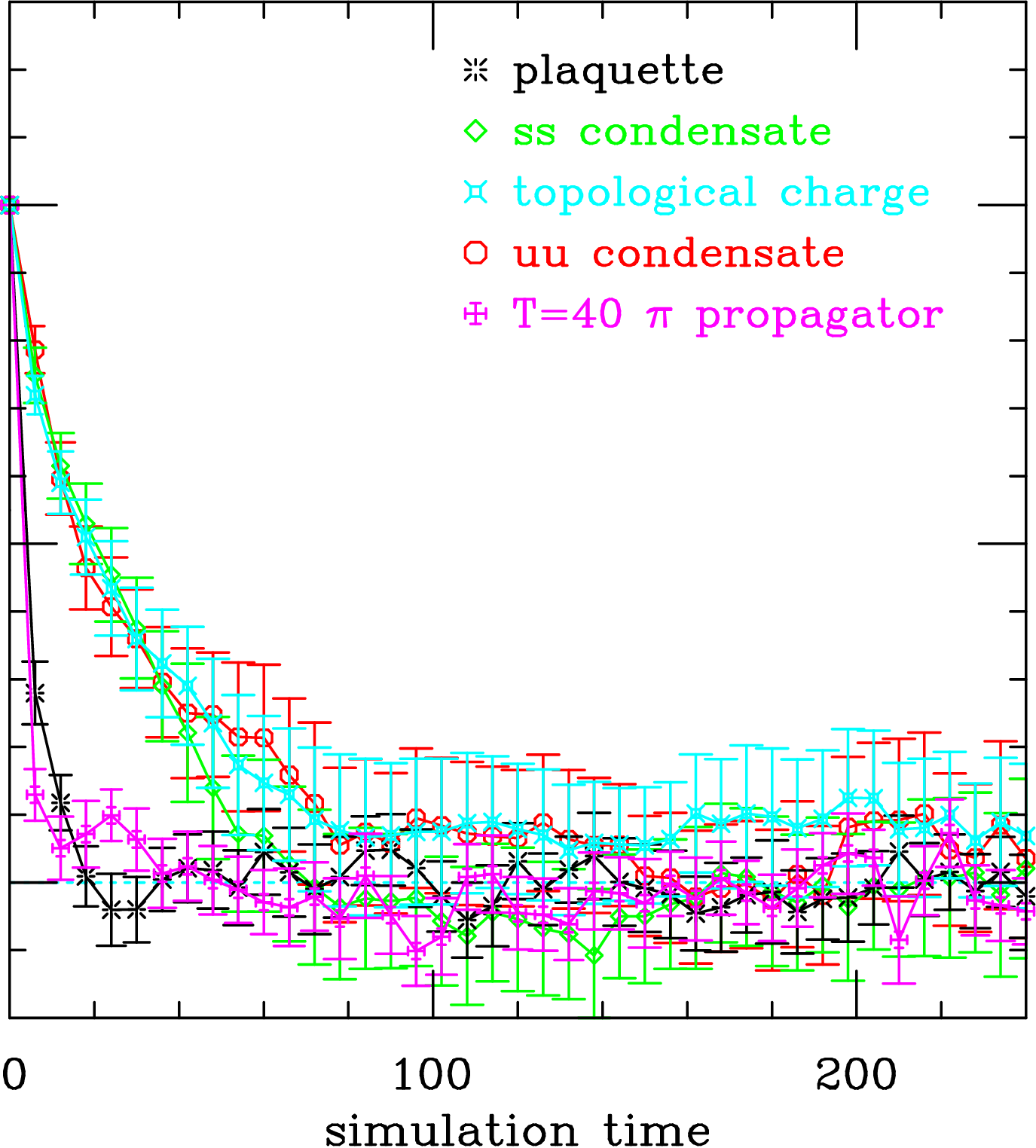}
\caption{Autocorrelations of various quantities in simulation time on the 
ensembles with $m_l=m_s/5$ and $a \approx 0.06$ fm (left) and $a \approx 0.09$ fm (right).}
\label{ac-vs-quantity}
\end{figure}

Fortunately, on many ensembles, we have multiple measurements using different 
stochastic sources for the condensate. This enables us to disentangle the 
two sources of fluctuation, by instead defining 

\begin{equation}
C(\mathcal{O};t \neq 0) = \left. 
\frac{\LL \mathcal{O}_{i} \mathcal{O}_{j} \RR 
- \LL \mathcal{O}_{i} \RR \LL \mathcal{O}_{j} \RR}
{\LL \mathcal{O}_a \mathcal{O}_b \RR - \LL \mathcal{O} \RR^2} 
\right|_{a \neq b, j=i+t}
\label{ac-def2}
\end{equation}
where $a$ and $b$ are different stochastic sources on the same configuration. 
(A similar redefinition of the numerator for the $t=0$ case recovers the 
result that $C(\mathcal{O};0) = 1$.)
This method obviously does not apply to ensembles where only a single 
stochastic source has been run on each configuration.

Figure~\ref{ac-vs-quantity} shows the autocorrelations of several observables 
in simulation time on two different ensembles. 
The errors on the autocorrelation at 
different separations are highly correlated.
We present results for the topological charge, the plaquette, the light and 
strange sea-quark condensates, and the 
(pseudoscalar) pion propagator at a separation corresponding roughly to the 
shortest length used in fits to determine $M_\pi$. 
On the finer of these ensembles ($a \approx 0.06$ fm), the topological charge has an extremely long autocorrelation length, followed by 
the strange- and light-quark condensates; on the coarser one ($a \approx 0.09$ fm), the autocorrelation length of 
the topological charge is comparable to that of the condensates. The autocorrelation length
for the plaquette and the pion propagator are quite short. Nonetheless, 
even for the pion propagator, there are non-negligible correlations
between the propagator on lattices separated by only a few time units, so 
some blocking procedure is necessary to correctly estimate the 
covariance matrix for the propagator when performing fits. 

\begin{figure}[t]
\centerline{
\includegraphics[height=0.468\textwidth]{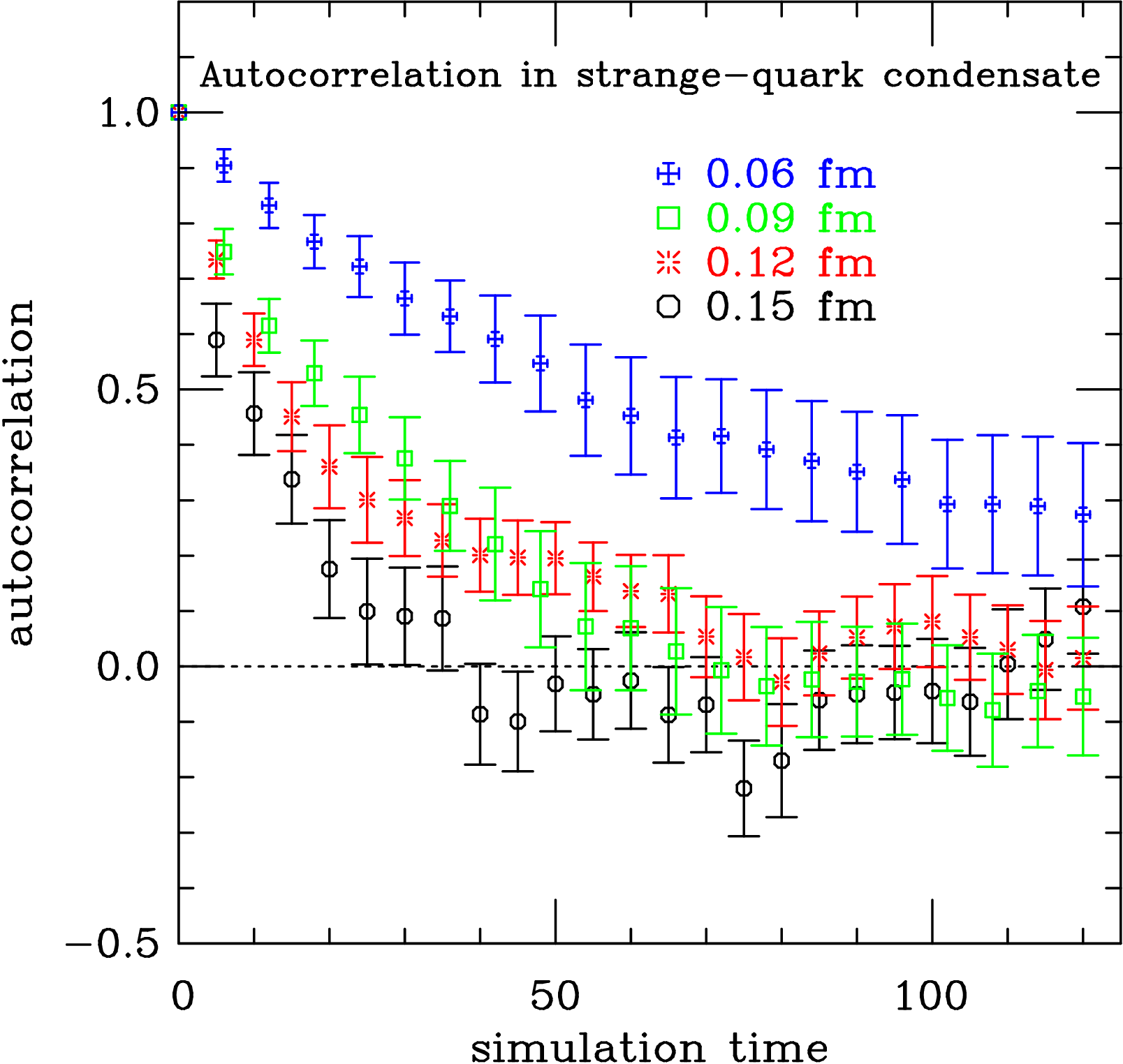}\hfil
\includegraphics[height=0.468\textwidth]{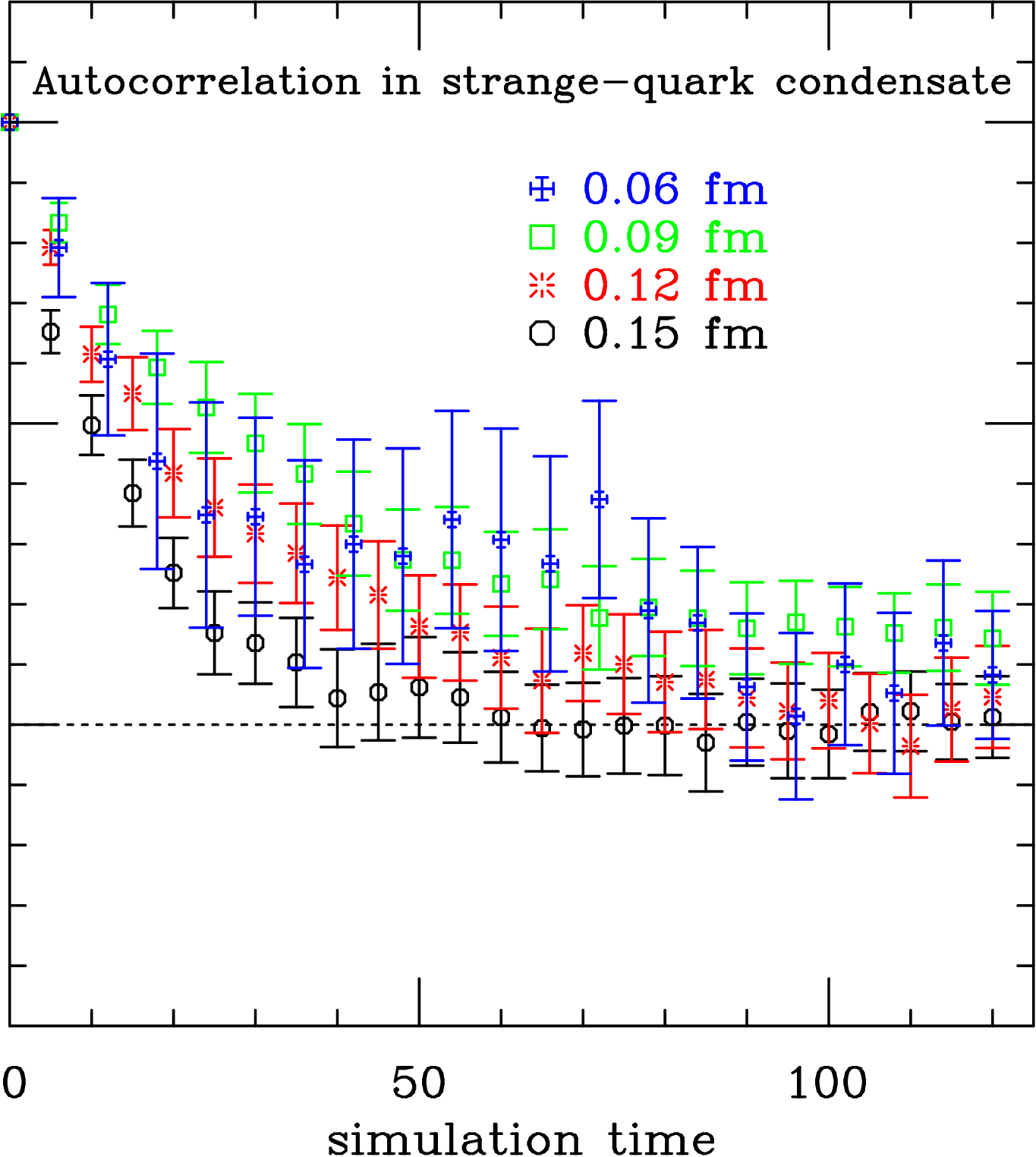}
}
\caption{Lattice spacing dependence of autocorrelations in the 
strange-quark condensate for light sea quark mass $m_l=m_s/5$ (left panel), 
and $m_l=m_s/10$ (right panel).
\label{ac-vs-a}
}
\end{figure}

We have three gauge ensembles which differ only in their spatial volume. 
They have $a \approx 0.12$~fm, $m_l = m_s/10$, and lattice volumes of 
$24^3 \times 64$, $32^3 \times 64$, and $40^3 \times 64$.
These ensembles provide us with an opportunity to examine whether the 
autocorrelation length depends on the lattice volume; we observe 
no such dependence. 

In Fig.~\ref{ac-vs-a} we show the autocorrelation function
for the strange-quark condensate at four lattice spacings with
light quark masses $m_l=m_s/5$ and $m_s/10$. The $m_l = m_s/10$ 
data suggest a trend toward increasing autocorrelation time with decreasing
lattice spacing, although the noisy $a \approx 0.06$ fm data do not fall
along that trend. Indeed, the $m_l = m_s/5$ data display this trend more 
strongly, with a greatly enhanced autocorrelation time for the finest ensemble.
The measurements required to do a similar comparison on the physical $m_l$ 
ensembles have not been completed. Topological charge data are available
for all of the $m_l = m_s/5$ ensembles, and a similar trend is apparent there, 
including a greatly enhanced autocorrelation time for the $a \approx 0.06$~fm,
$m_l=m_s/5$ ensemble.
It should be noted that the autocorrelation times for this ensemble, particularly for the
topological charge, are so long that they create difficulties in estimating the uncertainty 
in the autocorrelation function.

\begin{figure}[t]
\centerline{
\includegraphics[height=0.48\textwidth]{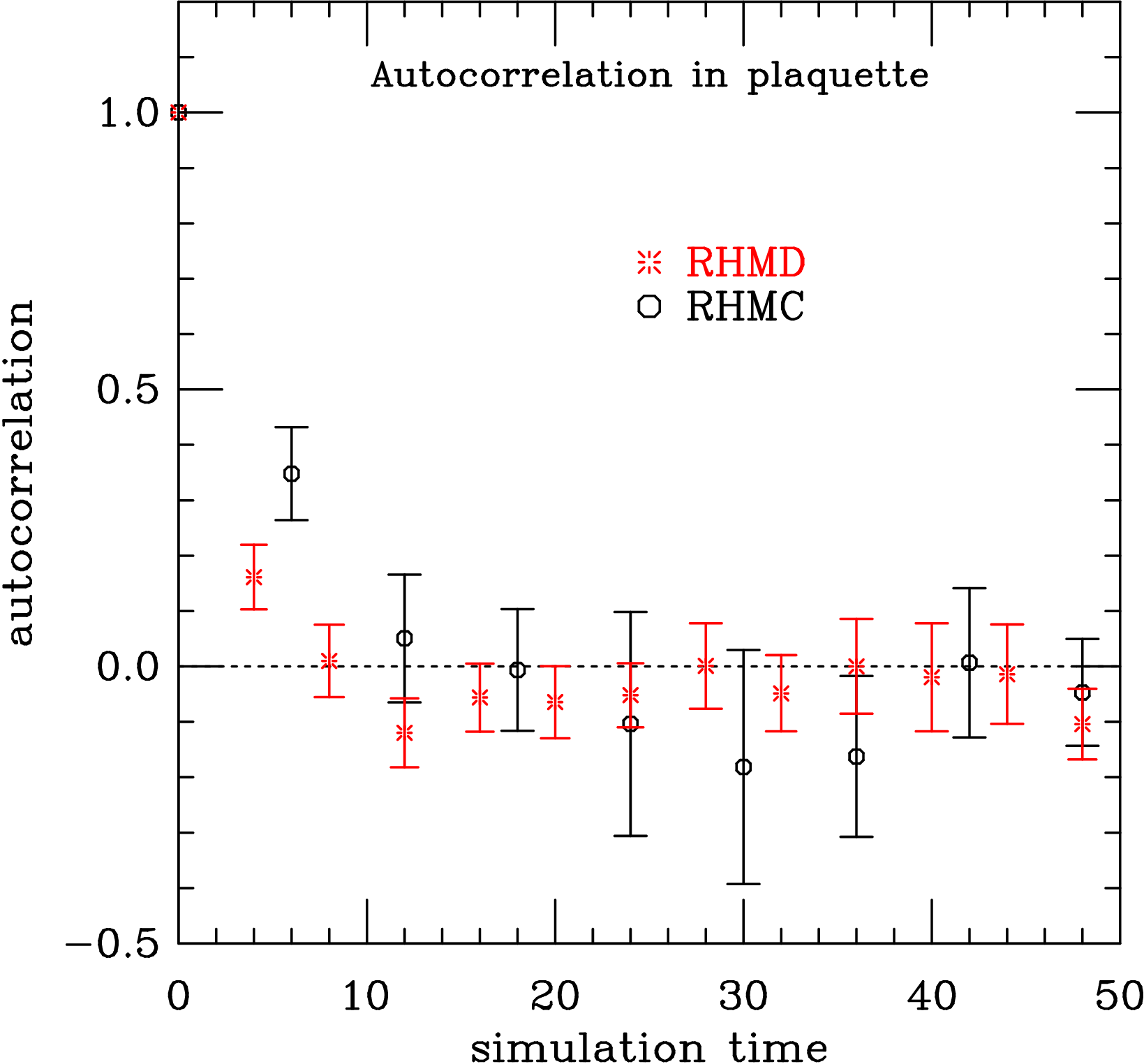}\hfil
\includegraphics[height=0.48\textwidth]{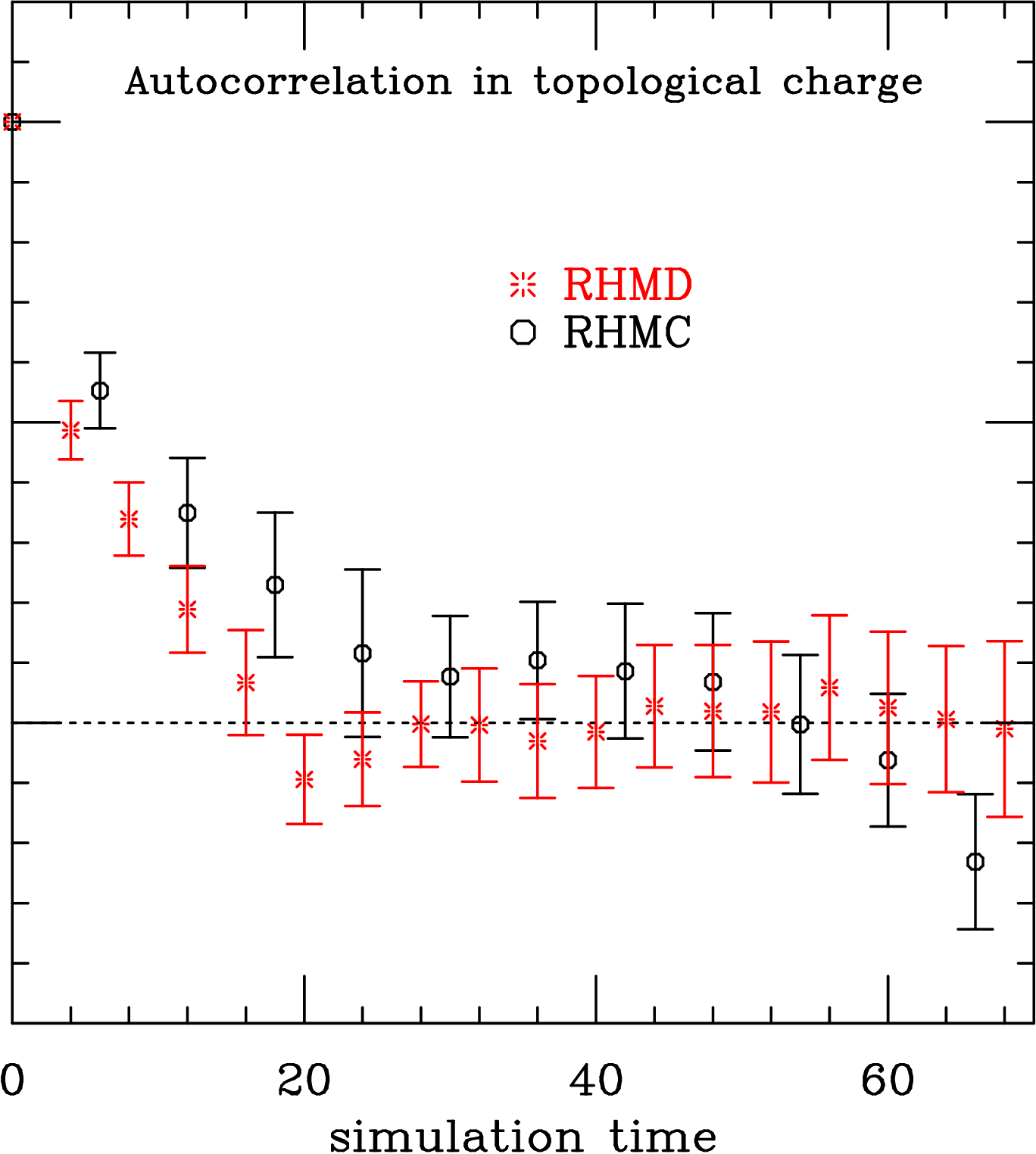}
}
\caption{Dependence of autocorrelation length on whether the Metropolis 
accept/reject step is present (RHMC) or absent (RHMD) in gauge generation,
on the ensemble with $a \approx 0.09$ fm, $m_l=m_s/27$. 
The left plot shows the plaquette, and the right plot shows the 
topological charge.
\label{ac-vs-algo}
}
\end{figure}

As discussed in Sec.~\ref{SIMULATIONS}, for large lattices
there are advantages to using the RHMD algorithm instead of RHMC. 
Use of RHMD will na\"{i}vely result in a decrease of
autocorrelation lengths by a factor equal to the Metropolis acceptance 
rate, generally about 70\%.  
To test for this effect, we compare autocorrelation
lengths on the RHMC and RHMD segments of the
$a \approx 0.09$ fm, physical light-quark mass ensemble which has roughly an
equal number of time steps with each algorithm. 
In anticipation of this decrease in the autocorrelation length
we have saved a gauge configuration every four time units in the 
RHMD evolution, under the assumption that this corresponds approximately
to every six time units using RHMC. Such a comparison is shown 
in Fig.~\ref{ac-vs-algo}. 
The expected decrease in the autocorrelation length is visible
in both the plaquette and the topological charge. It would be natural to
do a similar analysis for the strange-quark condensate,
as the quantity typically used to probe for thermalization, but those
data are too noisy on this ensemble to make any claim about the presence
or absence of the expected effect.

\section{Conclusions}
\label{CONCL}

We are nearing completion of the first phase of our project to generate
gauge configurations with four flavors of quarks with the HISQ action. 
In this phase
of our effort, we are generating ensembles with four values of the 
lattice spacing ranging from 0.06~fm to 0.15~fm, and three values of the 
light quark mass, including the value for which the Goldstone pion mass 
is equal to the physical pion mass. As can be seen from 
Tables~\ref{physical_strange} and \ref{unphysical_strange}, we have
completed all but three of the planned ensembles with $a\geq 0.06$~fm.
We hope to finish the remaining ones in the coming year. 
In the second phase of this project we plan additional ensembles at 
lattice spacings $a\approx 0.045$~fm and 0.03~fm. 
Most of the HISQ ensembles have been generated using the RHMC
algorithm, but we have found that for $a\leq 0.09$~fm, the gauge configurations
are smooth enough that one can use the RHMD algorithm, which results in
a large savings in computer resources. For $a\leq 0.09$~fm, the RHMD
algorithm introduces a systematic error, which is small compared to other
uncertainties, and, for all but a few quantities, is smaller than the
statistical error. Even for the worst case given in Table~\ref{tab:rhmdmasses},
the systematic difference between the RHMC and RHMD results is only
1.2 times the statistical error.

We have determined the lattice spacings of the ensembles by two different
methods. In one, we extract the Sommer parameter $r_1$ in lattice units
from the heavy-quark potential, and in the other we calculate the leptonic
decay constant $f_{p4s}$ of a fictitious pseudoscalar 
meson with valence quarks of mass $0.4\, m_s$, again in lattice units. 
Both of these quantities can be determined quickly
and accurately for a given ensemble. To obtain the lattice spacing in 
physical units, we must also determine $r_1$ or $f_{p4s}$
in physical units using one piece of experimental input.
Ultimately, this will be done on the HISQ lattices, but for the moment,
we take the physical results for $r_1$ and $f_{p4s}$ from
our more extensive data on the asqtad ensembles. We find excellent agreement
between the two approaches.

We have calculated the topological susceptibility on five of the HISQ ensembles,
including two with physical mass light quarks, and one with lattice spacing
$a\approx 0.06$~fm. 
We see from Fig.~\ref{fig:topo_hist}
that the simulation samples a wide range of values for the topological
charge, and does not appear to become stuck in any sector. The topological
susceptibility provides an excellent illustration of the improvement of
the HISQ ensembles relative to the asqtad ones, because it is calculated
without involving valence quarks. Figure~\ref{fig:HISQtopo} shows this
improvement, and also demonstrates the decrease in the susceptibility
with pion mass in accordance with the expectation from chiral
perturbation theory. 

Taste-symmetry breaking for pseudoscalar mesons is discussed in detail
in Section~\ref{TASTE}, and a variety of results for it are tabulated in
the appendix. We find that the $\bar ll$, $\bar ls$ and $\bar ss$
systems can be understood in terms of \schpt. They all exhibit
an accidental $SO(4)$ symmetry, but some small breaking of this symmetry
is present in the $\bar ss$ system at the coarsest lattice spacing. 
For the $\bar lc$ and $\bar sc$
systems, \schpt\ is not applicable, and the appropriate effective field
theory is ``all staggered heavy meson chiral perturbation theory,'' \aschpt.
$SO(4)$ symmetry is broken, but the pattern of taste splittings can be
understood using \aschpt. There is no chiral effective theory available to 
analyze the $\bar cc$ system. We see from the data that the breaking of $SO(4)$ 
symmetry is very strong for this case, but the relative taste splittings
are much smaller than in the light meson sector. Indeed, the difference
between the RMS pseudoscalar mass and the mass of the taste $\xi_5$
pseudoscalar (the Goldstone pion in the case of the $\bar{l}l$ system)
is 61~Mev, 26~MeV and 19~MeV for the $\bar{l}l$, $\bar{s}s$ and
$\bar{c}c$ systems, respectively on the $a\approx 0.12$~fm, $m_l=m_s/5$
ensemble. 

We save a gauge configuration every five molecular dynamics time units
for ensembles with $a > 0.09$~fm, and every six time units for ensembles
with $a\leq 0.09$~fm time units. Successive gauge configurations are, of
course, correlated, and these correlations must be taken into account
when analyzing the statistical uncertainties of measurements. The
autocorrelation length depends on the quantity being measured, the lattice
spacing and the light quark mass. A number of examples are given in
Sec.~\ref{AUTO}.

We plan to make these gauge ensembles publicly available, and we believe
that they will be useful for the study of a wide range of problems in
high energy and nuclear physics.

\vspace{0.2in}
\centerline{\bf Acknowledgments}
\vspace{0.2in}

Computations for this work were carried out with resources provided
by the USQCD Collaboration, the Argonne Leadership Computing Facility,
and the National Energy Research Scientific Computing Center, which are
funded by the Office of Science of the U.S. Department of Energy; and with
resources provided by the National Center for Atmospheric Research, the
National Center for Supercomputing Applications, the National Institute 
for Computational Sciences, and the Texas Advanced Computing
Center, which are funded through the National Science Foundation's
XSEDE Program. This work was supported in part by the U.S. Department of
Energy under Grants DE-FG02-91ER-40628, DE-FG02-91ER-40661, DE-FG02-04ER-41298,
DE-FC02-06ER41446, and DE-FC02-06ER-41439; and by the 
National Science Foundation under Grants
PHY07-57333, PHY07-03296, PHY07-57035, PHY07-04171,  PHY09-03571, PHY09-70137,
PHY09-03536, and PHY10-67881.
This manuscript has been co-authored by an employee of Brookhaven Science 
Associates, LLC, under Contract No. DE-AC02-98CH10886 with the U.S. 
Department of Energy. 
Fermilab is operated by Fermi Research Alliance, LLC, under Contract No. 
DE-AC02-07CH11359 with the U.S. Department of Energy.

\newpage
\appendix

\section{Taste splitting of pseudoscalar mesons}
\label{app:taste_split}

In Tables~\ref{tab:taste_degenerate} and \ref{tab:taste_heavylight},
we tabulate the taste splittings of the pseudoscalar
mesons made from combinations of light, strange and charm quarks.

\begin{table}[h]
\begin{tabular}{|c|c|c|c|c|c|c|c|}
\hline
$\approx a$~(fm) &   $\gamma_0\gamma_5$ & $\gamma_i\gamma_5$ & $\gamma_i\gamma_0$ & $\gamma_i\gamma_j$ & $\gamma_0$ & $\gamma_i$ & ${\bf 1}$ \\
\hline
\multicolumn{8}{|c|}{light-light} \\
\hline
0.15   & 0.1000(71j)& 0.1033(8j) & 0.1959(29j)& 0.1901(57j)& 0.2855(46j)& 0.2872(114j)& 0.3678(441j) \\
\hline
0.12   & 0.0553(28s)& 0.0542(11s)& 0.1031(21s)& 0.1013(48s)& 0.1516(32s)& 0.1616(72s)& 0.2068(172s) \\
\hline
0.09   & 0.0180(19j)& 0.0180(5j) & 0.0335(10j)& 0.0328(26j)& 0.0493(14j)& 0.0485(32j)& 0.0631(51j) \\
\hline
0.06  & 0.0031(5j) & 0.0033(1j) & 0.0062(2j) & 0.0064(6j) & 0.0090(3j) & 0.0093(7j) & 0.0128(7j) \\
\hline
\multicolumn{8}{|c|}{strange-strange} \\
\hline
0.15   & 0.0948(14s) & 0.1009(12s) & 0.1853(19s) & 0.1806(19s) & 0.2626(24s) & 0.2504(29s) & 0.3186(50s) \\
\hline
0.12   & 0.0464(23s) & 0.0486(15s) & 0.0932(17s) & 0.0897(23s) & 0.1320(23s) & 0.1303(23s) & 0.1674(36s) \\
\hline
0.09   & 0.0212(47s) & 0.0157(39s) & 0.0283(40s) & 0.0315(47s) & 0.0386(48s) & 0.0458(48s) & 0.0601(56s) \\
\hline
0.06   & 0.0028(10j) & 0.0027(1j) & 0.0049(4j) & 0.0059(11s) & 0.0071(4j) & 0.0071(5j) & 0.0092(6j) \\
\hline
\multicolumn{8}{|c|}{charm-charm} \\
\hline
0.15   & 0.3229(25s) & 0.5432(27s) & 0.7847(33s) & 0.6354(27s) & 0.9431(37s) & 0.8224(33s) & 0.9617(41s) \\
\hline
0.12   & 0.1394(34s) & 0.2227(34s) & 0.3446(39s) & 0.2754(36s) & 0.4324(41s) & 0.3706(39s) & 0.4485(44s) \\
\hline
0.09   & 0.0486(78s) & 0.0719(78s) & 0.1129(85s) & 0.1000(74s) & 0.1410(85s) & 0.1355(85s) & 0.1748(91s) \\
\hline
0.06  & 0.0143(21s) & 0.0189(3j) & 0.0317(6j) & 0.0275(21s) & 0.0423(7j) & 0.0376(6j) & 0.0462(7j) \\
\hline
\end{tabular}
\vspace{0.1in}
\caption{\label{tab:taste_degenerate}
Taste splittings for pseudoscalars  with equal valence-quark masses on 
the $m_s/5$ ensembles.
These are the splittings for the pseudoscalar mesons with valence quark masses 
equal to the light sea quark mass, $m_s/5$, equal to the sea strange quark mass,
and equal to the sea charm quark mass.
For each non-Goldstone taste $\xi$ we tabulate the squared mass
difference $r_1^2 a^2 \Delta_{\xi}= r_1^2 (M^2_{\xi}-M^2_{5})$,
where the errors are statistical only.
”j” indicates the
error comes from a jackknife analysis, and “s” indicates the error is 
just the error from the mass of the non-Goldstone pion.  This is a 
reasonable approximation of the true statistical error in the
squared-mass splitting when either (i) the uncertainty in the Goldstone 
pion mass is much smaller than
that of the non-Goldstone pion, and/or (ii) the Goldstone and non-Goldstone 
pion masses are strongly correlated.
}
\end{table}

\begin{table}
\begin{tabular}{|c|c|c|c|c|c|c|c|c|}
\hline
\multicolumn{9}{|c|}{\phantom{XXXXXXXXX}light-charm} \\
\hline
                      & $\gamma_5$  & $\gamma_0\gamma_5$ & $\gamma_i\gamma_5$ & $\gamma_i\gamma_0$ & $\gamma_i\gamma_j$ & $\gamma_0$ & $\gamma_i$ & ${\bf 1}$ \\
\hline
$\gamma_5$            & --          &  0.088(17)  & 0.118(17)   & 0.164(23)   & 0.137(19)   & 0.205(29)   & 0.151(28)   & 0.185(31) \\
\hline
$\gamma_0\gamma_5$    & -0.085(4) &  --         & {\bf \R{0.031(20)}}   & 0.076(26)   & 0.049(19)   & 0.117(30)   & 0.063(31)   & 0.097(33) \\
\hline
$\gamma_i\gamma_5$    & -0.126(4) & {\bf\R{-0.041(4)}}  & --          & 0.045(23)   & 0.019(20)   & 0.086(30)   & 0.033(29)   & 0.066(35) \\
\hline
$\gamma_i\gamma_0$    & -0.205(6) & -0.120(7) & -0.079(6) & --          & {\bf\R{-0.027(25)}}  & 0.041(26)   & -0.013(24)  & 0.021(29) \\
\hline
$\gamma_i\gamma_j$    & -0.174(5) & -0.088(4) & -0.048(5) & {\bf \R{0.031(8)}}  & --          & 0.068(37)   & 0.014(33)   & 0.048(35) \\
\hline
$\gamma_0$            & -0.266(8) & -0.181(8) & -0.140(7) & -0.061(5) & -0.092(10) & --          & {\bf\R{-0.054(29)}}  & -0.020(26) \\
\hline
$\gamma_i$            & -0.229(7) & -0.144(7) & -0.103(6) & -0.025(5) & -0.056(8) & {\bf \R{0.036(6)}}  & --          & 0.034(33) \\
\hline
${\bf 1}$             & -0.290(8) & -0.205(8) & -0.164(8) & -0.085(6) & -0.116(9) & -0.024(7) & -0.060(6) & --        \\
\hline
 \multicolumn{9}{|c|}{strange-charm\phantom{XXXXXXXXX}}  \\
\hline
\end{tabular}
\caption{\label{tab:taste_heavylight}
Taste splittings for the heavy-light pseudoscalars on the $a\approx 0.15$ fm 
$m_s/5$ ensemble.
Since the breaking of the approximate $SO(4)$ symmetry is important here, 
and all the masses
are correlated so naive combination of errors is incorrect, we 
tabulate all of the splittings.
As in Table~\protect\ref{tab:taste_degenerate}, we tabulate 
$r_1^2 \LP M_{\xi_2}^2-M_{\xi_1}^2 \RP$,
where $\xi_2$ and $\xi_1$ are the tastes in the row and column labels.   
Since this matrix is
antisymmetric, we have placed the light-charm splittings in the upper 
triangle, and the
strange-charm splittings in the lower triangle.   All errors are estimated 
using a jackknife procedure.
The $SO(4)$ breaking splittings are shown in {\bf \R{red}}.
}
\end{table}

\end{document}